# Glass theory: ground and excited states of coupled electron pairs

Jia Lin Wu

*College of Material Science and Engineering, Donghua University, Shanghai, 201620, China,*

Since the discovery of the strict second-order-delta-vector (magnetic moment) (SODV) theory of Gennes $n = 0$, the theoretical community has been searching for SODVs that can evolve from complex glass states to biomolecular systems. In the theoretical study of the abnormal viscosity of entangled polymer melts, we unexpectedly found an SODV. It is a synchronous-antisymmetric coupled electron pair (CEP) excited state that creates a dynamic interface between two slightly overlapping adjacent hard-sphere molecules (HSMs). The two HSMs suddenly acquired the identical new spin in opposite directions, so the two-dimensional soft matrix predicted by de Gennes was found in the glass model. Unlike electronic excited states, the energy of CEP excited states is three orders of magnitude smaller than that of electronic excited states, and they appear in the form of a nano-scale dynamic Ising models. This new mathematical physics regime can directly explain almost all glass and glass transition phenomena. In this paper, two paradigms with $n = 0$ are given, and discussed its wide application prospects.

Subject Areas: Glass Transition, Molecular Dynamics, Soft Matter, Coupled Electron Pair

.

## I. INTRODUCTION

About half a century ago, de Gennes $n = 0$ SODV theory [1] was discovered and caused a sensation in the theoretical world. However, in the next 40 years, since the emergence and application of SODV have not yet been discovered, the enthusiasm for $n = 0$ has gradually subsided [2]. The strict $n = 0$ theory is a mean field theory. Regardless of the complexity of the molecular chemical structure, interacting with countless molecules, each molecule from solid to liquid is an HSM, vibrating along the $q$-axis in the Lennard–Jones (L–J) potential field [3]. Therefore, one of the challenges of glass theory is what is the interaction between mean field HSMs? How are HSMs clustered? How does HSM move? What is the disordered rigid mechanism of HSM glass？ What is the cause of the broad relaxation time spectrum of the HSM model? These five questions can be reduced to one question: which theory can answer all five questions? One of the weaknesses of current glass theories is that mathematical models based on beauty capture some of the most intriguing features of glassy behavior but are too unrealistic to provide bases for predictive the interaction between HSMs [4]. The central assumption of the glass model proposed by de Gennes (the founder of soft matter theory) is that there is a low-density *soft matrix* in the HSM model [5]. This prediction by de Gennes may be based on his deep thinking on the $n = 0$ theory, and his profound insights into the shortcomings of the *no neighborhood effect* [6] in spin glass theory and neglecting geometric frustration [5] in mode coupling theory [7, 8].

The neighborhood effect mentioned here means that the spin interaction occurs only between two adjacent HSMs. This means that one of the core concepts of soft matter is the soft matrix to be explored. The key to the de Gennes glass model is how spin interactions occur between HSMs?

The 3.4 power law [9-11] of entangled polymer melt viscosity is sensitive to changes in the HSM theory. So far, no molecular theory has predicted the 3.4 power law. Thus, a distinctive theoretical approach of exploring molecular clustering and movement is adopted. First, the general expression for predicting melt viscosity using the five-HSM/five-cluster/five-local field model [12, 13] (Appendix A.6) is: $\eta \sim N^{\,9\,(1-Tg/Tm)}$, which is highly consistent with the experimental data of all known flexible and non-flexible polymers (Table 1). Furthermore, based on this expression, many previously unknown and amazing HSM clustering and movement attributes are derived as shown below. (i) The left-right asymmetry of the L−J potential and the five-HSM clustering fixed point position can define the (spin) orientation of the central HSM $a_0$. For the nine fixed points of the nine L − J potentials (A.4), at a fixed point from $t_0$ to $t_8$, four adjacent HSMs ($b_0$, $c_0$, $d_0$ and $e_0$) from different

Email: jlwu@dhu.edu.cn

directions are projected one after another on the $z$-axis ($q$-axis in Fig. 1) of their center HSM $a_0$ and are subsequently coupled and clustered. Thus, the $z$-axial two-dimensional (2D) cluster $z$-$V_i(a_0)$ (with relaxation time $\tau_i$) and the three-dimensional (3D) hard sphere $z$-$\sigma_i(a_0)$ (where $i$ = 0, 1, 2 ... 8), centered on $a_0$ are derived in the order of increasing size (A.2). (ii) Since the 200-HSM is the "critical molecular weight" obtained through experiments, at the ninth fixed point, it can be inferred that the 200-chain-HSM $z$-$V_8(a_0)$ (with a $\tau_8$ time-scale) around $a_0$ is the $z$-axis 2D soft matrix (Fig. 9). (iii) The Hamiltonian $H$ is the emerging energy to form a soft matrix, a material parameter suitable for the entire temperature range from solid to liquid. An increase in temperature is always accompanied by an increase in the number of soft matrices, and vice versa. $H = k_B T_g°$ in a $\mu$-direction soft matrix, which is the largest ordered structural unit against thermal fluctuation in the $\mu$-direction; the average energy of all soft matrices randomly oriented in a system with temperature $T$ is still $k_B T_g° = H = k_B T_g$. Additionally, the energy of the rearranged soft matrix is $k_B T_m° = k_B T_m$. The rearranged soft matrix is actually the sequential projection of the four soft matrices of four adjacent HSMs on the $-z$-axis completely canceling the $a_0$ soft matrix, and its energy $k_B T_m° = k_B T_g° + 4\varepsilon_0$ (Fig. 5) for flexible polymers. Here, $k_B T_g°$ and $k_B T_m°$ represent two ordered energies associated with the soft matrix, which can be represented by two disordered energies that are in equilibrium with the two ordered energies, independent of the temperature $T$ of the system. (iv) In the reptation tube model proposed by de Gennes and perfected by Doi and Edwards [14,15], the chain of length $N$ must be replaced by a "completely free diffusion chain" consisting of $N^*$ equivalent particles with $N^*$ degrees of freedom (DoFs). $N^* = N_x^* \cdot N_y^* \cdot N_z^* := N^{3(1-T_g/T_m)}$, where $N_x^*$, $N_y^*$ and $N_z^*$ are the number of DoF required for the chain $N$ to jump the $n_z$ ($\le 0.036$, A.5), less than the HSM vibration amplitude $\sim 0.1$) steps along the $x$-, $y$-, and $z$-axes, respectively. In popular statistical theory $N^* \sim N$, does not match the experimental results, and the most basic relationship of the moving unit length $l$ in 3D space: $l^2 = l_x^2 + l_y^2 + l_z^2$ is invalid here, HSM walking follows new statistical law in Figs. 3(e) and 4(d) derived from the $n = 0$ theory, and never appears in existing literature [16-21].

The above four inferences are based on the assumption that there are two synchro-orthogonally coupled electrons on the interface between two adjacent HSMs, which are transported in parallel from one end of the interface to the other, called the interface excitation (IE) [12,13]. Although the origin of IE is unclear, it is impossible to prove the 3.4 power law without introducing the concept of IE (A.6). Therefore, this research approach leads to a clear goal: the correlation between $n = 0$ SODV and IE must be found.

## II. RESULTS AND DISCUSSION

**Dynamic 2$\Delta$d microcubic lattice inside each HSM.** In the nine L–J potential fields, the clustering positions of two adjacent HSMs cannot be arbitrarily, but is controlled by the Lindemann ratio $d_L = (q_{iR} - q_{iL})/\sigma_i = (q_R - q_L)/\sigma = 0.1046...$ (A.4, Eq. 9), which describes the overall thermal fluctuation. When two HSMs are coupled into a cluster, each HSM centroid [also the positive charge-center particle of HSM ($M^+$-P)] can only be located at a fixed point in its own field, or on a plane away from its vibration balance position $\Delta$d (Fig. 1). Thus, during the clustering of four adjacent HSMs in sequence, the trajectory of $M^+$-P of each HSM will outline a microcubic lattice with 2$\Delta$d sides and centered on its vibration equilibrium position. $\Delta$d = $(q_{iR} - q_{i0})/\sigma_i = (q_R - q_0)/\sigma \approx 0.055$ (A.4, Eq. 10). $(q_{iR} - q_{i0})/\sigma_i = (q_R - q_0)/\sigma$ is a scale transformation, meaning that the clustering graph of the nine clusters $z$-$V_i(a_0)$ in Fig. 1 only needs to be represented by one $z$-$V_0(a_0)$ in Figs 3 or 4 [when $m$ ($m = i + 1$) IE-closed loops appear around $a_0$, the four adjacent HSMs of $a_0$ must have completed $m-1$ closed loops. By analogy, $z$-$V_i(a_0)$ can be obtained, that is, HSM IE spin $z$-$S_m(a_0)$ is equivalent to cluster $z$–$V_i(a_0)$, (Figs 6–9)].

**Dynamic cubic lattice of HSM.** Within the relaxation time $\tau_0$, HSM $a_0$ has a dynamic cubic lattice (HSCL) with $(1+ d_L)$ sides generated by its four IE new interfaces in sequence. The dynamic $u$ interface is actually $k_u$ transient excited states of the $z$-axis CEP, which appear in turn in a 90° $u$-space between two adjacent HSMs from one end to the other [Figs 2(b), 3(b) and 4(a)], $k_u$ excited states are also called $k_u$ directed repulsive electron pairs (DREPs), $u \in \alpha, \beta, \gamma, \delta$, they are the four new interfaces around HSCL. The equilibrium forms of attraction and repulsion of two HSMs in the clustering, such as $a_0$ and $c_0$, are shown in Figs 3 and 4. The center of the HSCL is the vibration equilibrium center at the bottom of the potential well. The HSM has a $2\Delta d$ micro-cubic lattice (instead of $2\Delta d$ micro-spheres) means that the walking of HSM is only along the local $\pm x$, $\pm y$ and $\pm z$ axis. Each face of the micro-cubic lattice is an equipotential with energy $f_c$ in Fig. 1, where the unit 1 used to measure $\Delta d$ and the unit 1 used to measure $\sigma$ are both measured by the potential well $\varepsilon_0 = 1$. Since $2\Delta d > d_L$, two adjacent HSMs $a_0$–$c_0$ overlap at the $\beta$-interface [the plane of $y = 1/2 (1 + d_L) \approx 0.5023$] on the $x$-$y$ projection plane, thus generating a new attracting potential balanced with the repulsive energy of the $k_\beta$ electron unclosed orbit pairs of the $k_\beta$ excited states of CEP. Thus, $k_\beta$ DREPs appear sequentially from point 2 to point 3 on the $\beta$-interface [Figs 3(c) and 4(c)], which is the theoretical origin of the previously predicted $\beta$-IE arrow [12,13]. The physical image of the excited states of CEPs are: $k_\alpha + k_\beta + k_\gamma + k_\delta$ DREPs generated when the four HSMs are coupled with the central HSM $a_0$ make $a_0$ get a IE spin; once $z$-$S_9(a_0)$ (the $z$-axis DREPs form nine closed loops around $a_0$) appears, or once a $z$-soft matrix of 200 HSMs around the center HSM $a_0$ appears, the cavity with potential well energy $\varepsilon_0$ will appear at the $a_0$ position and carry 200 HSMs with 320 IE states, jumping $n_z$ ($\leq \Delta d$) steps along the $z$-direction. That is, the $z$-axis movement of an HSM is the jump of the $z$-axis IE spin system (soft matrix) of the HSM.

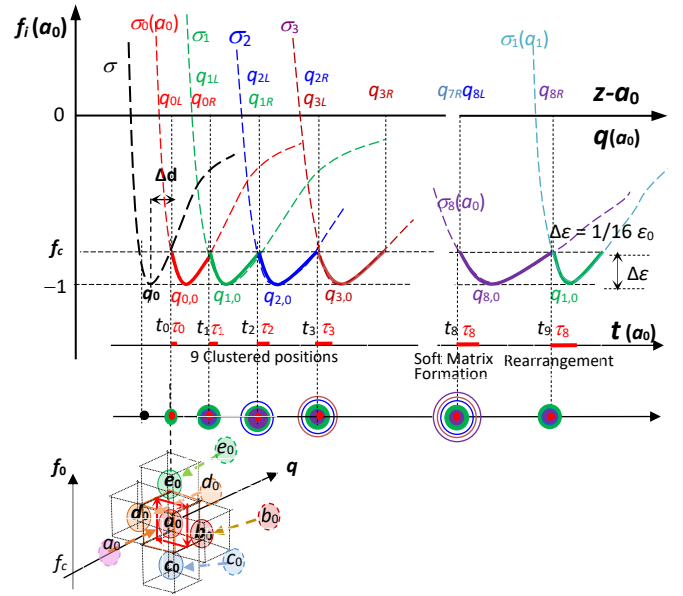

FIG. 1. Absence of random thermal vibrations in five-HSM clustering at the fixed points. When the randomly vibrating HSM $a_0$ (dotted ball) moves to the first fixed point ($f_c$, $q_R$), $q_R = q_{0L}$, four adjacent HSMs $b_0$, $c_0$, $d_0$, and $e_0$ (four dotted balls) in directions different from the $a_0$ direction also sequentially move to the fixed point. Within the relaxation time $\tau_0$, the four red arrows on the four successively excited interfaces of HSM $a_0$ (each arrow is a parallel jump-transport of CEP excited states from one end to the other on this interface, as shown in Fig. 3(c) and Fig. 4(c), form a "magnetic moment". Therefore, only during this time, HSM $a_0$ has a dynamic HSCL and obtains the $z$-axis HSM IE spin, labeled $z$-$S_1(a_0)$. As a cost of clustering, $a_0$ and each of its four adjacent HSMs simultaneously lose a pair of random thermal vibrations from $q_R$ back to $q_0$.

**The $\vartheta_\lambda$-state in random thermal vibration inside HSM.** The L– J potential ignores all synchro-antisymmetrically coupled (SASC) $\vartheta_\lambda$-$\vartheta_{\lambda*}$ states that occur sequentially in every two adjacent HSMs in the cluster [Fig. 2(a)]. $\lambda$ is a number and $\vartheta_\lambda$ refers to the $\lambda$th spatial angle-line state numbered according to the order of appearance from one end trap state to the other on the $u$ interface. Let the connecting line of two tangent points ($p$ and $p_e$) where the electron orbit of the instantaneous position of hydrogen atom positive charge-center particle ($H^+$-P) is tangent to two parallel lines be written as the angle-line vector $\mathbf{V}_{\lambda u}(a_0)$ in $a_0$ HSCL. The subscript $\lambda_u$ is a set of numbers that appears on the interface $u$ between the two HSMs in sequence according to the order of the numbers: $\lambda_u = 1, 2, 3…\lambda, \lambda+1…k_u$ (the italicized numbers in this article indicate instantaneous states). $\mathbf{V}_\lambda$-$\mathbf{V}_{\lambda*}$ is the two position angle-lines of two SASC $M^+$-Ps that can be clustered. At the two pairs of coupling points $p$–$p^*$ and $p_e$–$p_e{}^*$ on these two vectors, the two electrons in the two HSMs are balanced with their positive charges before and after clustering, respectively, (defined as the $\vartheta_\lambda$-$\vartheta_{\lambda*}$ state of two adjacent HSMs), which contains four features. (i) HSM vibration equilibrium center and $M^+$-P and $H^+$-P and point $p_e$ are located on $\mathbf{V}_\lambda$-angle-line, and satisfy: $\vartheta_\lambda(a_0) \cdot \vartheta_{\lambda*}(c_0) = \delta_{\lambda\lambda*}$ (SODV), which

Indicates that the two HSMs are now at a transient steady ground state (GS). (ii) One parallel line (tangent point $p$) is a diagonal line on the $u$ interface and the other (tangent point $p_e$) is the vibration direction specifically selected by the electron in its C–H bond resonance (labeled $\vartheta_\lambda$-GS in Fig. 2(a). (iii) Both $M^+$-P and $H^+$-P ($H^+$-P jumps perpendicular to the diagonal plane) lying on the same diagonal plane simultaneously perform a jump parallel to the diagonal and deflect a tiny angle $\varphi_\lambda$ from $\vartheta_\lambda$-angle-line to $\vartheta_{\lambda+1}$-angle-line. (iv) Since the electron is fast moving, when $M^+$-P and $H^+$-P deflect $\varphi_\lambda$ angle, the electron escapes the $\vartheta_\lambda$-GS and make a parallel transport around its H$^+$-P to complete the $(2\pi + \varphi_\lambda)$ in Fig. 3(b) [or $(2\pi - \varphi_\lambda)$ in Fig. 4(a)] non-closed loop, called the $\varphi_\lambda$-unclosed orbit. Where $\varphi_\lambda$ the deflection is angle of an electron on the $\vartheta_\lambda$-unclosed orbit from $\vartheta_\lambda$-GS to $\vartheta_{\lambda+1}$ GS, and $\varphi_\lambda$ is also the jumping angle of CEP from the $\vartheta_\lambda$-angle-line to the $\vartheta_{\lambda+1}$-angle-line. When M$^+$-P jumps and deflects $k_\beta$ times and jumps from point 10 to point 11 in Fig. 4(a) [or in Fig. 3(b)] on $\beta$-interface, the electron in the $k_\beta$ $\varphi_\lambda$-unclosed orbits and its DREP on $\beta$-interface deflects a total of $\pi/2$ angles around M$^+$-P. Both $k_u$ and $\varphi_\lambda$ are material parameters, it can be predicted that the $\varphi_\lambda$-unclosed orbit can carry chemical structure information of M$^+$-P and H$^+$-P, thus making the soft matrix of each material have different $kT_g$.

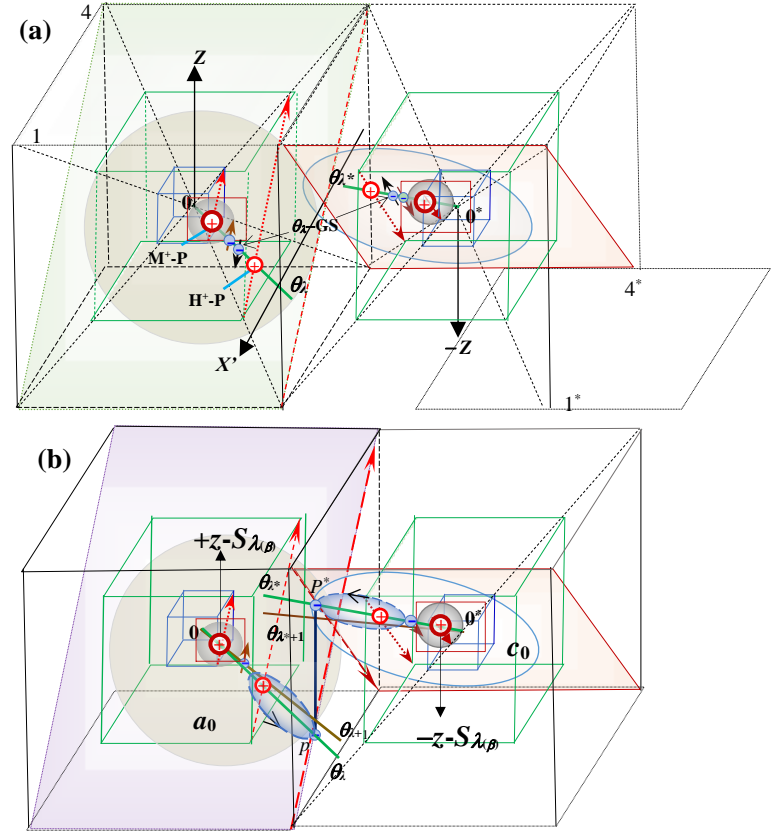

FIG. 2. Second-order $\delta_{\lambda\lambda*}$ vector (magnetic moment) of $n = 0$. (a) In clustering, when the HSCL $a_0$ (the vertices 1 and 4) are turned 180° around the $x'$-axis to become the HSCL $c_0$ (the vertices $1^*$ and $4^*$), it means that $a_0$ and $c_0$ are antisymmetric with respect to the $x'$-axis. HSM vibration equilibrium center 0, and $M^+$-P and $H^+$-P are located on $\vartheta_\lambda$-angle-line, which is a delta vector $\mathbf{V}_\lambda$ in fluctuation. When the vibration directions (two blue arrows) of two electrons in two C–H bonds on two orthogonal diagonal plans are parallel to their respective diagonal lines, the CEP at this time is $\vartheta_\lambda$-GS. (b) When the two SASC $M^+$-Ps (and $H^+$-Ps) jump from $\vartheta_\lambda$ to $\vartheta_{\lambda+1}$, the "eigenvalues" that make the directions of the two SASC vectors unchanged are the two points $p$ and $p^*$ where the two SASC $\varphi_\lambda$ unclosed orbits intersect on the two vectors $p$ and $p^*$, also satisfy: $\mathbf{V}_\lambda(p) \cdot \mathbf{V}_{\lambda*}(p^*) = \delta_{\lambda\lambda*}$. And these two SASC electrons are actually a CEP on the interface. Because their repulsive energy and the "connected orientation" of the two positions also connect the two $\varphi_\lambda$-$\varphi_{\lambda*}$ unclosed orbits of the CEP, the $\lambda$-th excited state of the $z$-axis CEP is the second-order delta magnetic moment with two SASC $\varphi_\lambda$ unclosed orbits.

**The $\vartheta_\lambda$-operation element in soft matrix.** Definition: A "deflection-jump of $\varphi_\lambda$ angle" in the random process from $\vartheta_\lambda$-GS to $\vartheta_{\lambda+1}$-GS is a $\vartheta_\lambda$-operation element in soft matrix $\mathcal{A}$. In geometry, the eigenvector **v** and the eigenvalue **λ** have a relation with the matrix **A**: **Av** = **λv**. The direction of the vector **v** after **A** transformation is unchanged, and only stretching or shortening are performed. In our study, we discuss two SASC positive charge angle-lines (M$^+$-Ps and H$^+$-Ps) of two adjacent HSMs and their two SASC electrons jump synchronously from $\vartheta_\lambda$-state to $\vartheta_{\lambda+1}$-state. In fluctuations, two SASC delta vectors in two adjacent HSMs $a_0 - c_0$ satisfy: $\mathbf{V}_\lambda(a_0) \cdot \mathbf{V}_{\lambda*}(c_0) = \delta_{\lambda\lambda*}$ (SODV). When M$^+$-P and H$^+$-P deflect $\varphi_\lambda$ angle, the electron in $\vartheta_\lambda$-GS can only form a $\varphi_\lambda$- non-closed orbit surrounding H$^+$-P in the opposite direction of H$^+$-P moving, from the point $p_e$ in $\vartheta_\lambda$-GS to that of $\vartheta_{\lambda+1}$-GS. The directions of both SASC $\mathbf{V}_\lambda$ and $\mathbf{V}_{\lambda*}$ vectors are unchanged. Each vector $\mathbf{V}_\lambda$ is stretched to the point $p$ where the $\varphi_\lambda$-non-closed orbit intersects at $\mathbf{V}_\lambda$, and $p - p^*$ forms a $\lambda$th-DREP in $\beta$-interface between $a_0$ and $c_0$. The eigenvectors are SASC $\mathbf{V}_\lambda(a_0) - \mathbf{V}_\lambda(c_0)$. The eigenvalue that does not change the directions of $\mathbf{V}_\lambda(a_0)$ and $\mathbf{V}_\lambda(c_0)$ is two "positions" where the two unclosed electron orbits are tangent to the interface, and they are located at two points on the two stretch delta vectors. The singularity is that the connecting line of the two points of $p - p^*$ makes the two HSMs suddenly obtain two antiparallel identical $z$-axis $n = 0$ spins after two coupled $\vartheta_\lambda$-operation elements: two coupled $\vartheta_\lambda$ and $\vartheta_{\lambda*}$ jump produce a $+z$-axis zero-spin $S_\lambda(a_0)$ and a $-z$-axis zero-spin $S_\lambda(c_0)$.

**The graph of $n = 0$ CEP.** In the vector model of de Gennes $n = 0$, the dimension of the orientation space $\Omega$ is $n$ [1], $\Omega$ is a phase space containing all orientations of spin $S_\lambda(a_0)$ at $\lambda$th-DREP. Here, the orientation space $\Omega$ contains only an *instant* $\delta_{\lambda\lambda*}$ direction, which is an instantaneous $z$-direction state of two electrons moving on two SASC $\varphi_\lambda - \varphi_{\lambda*}$-unclosed orbits, and its spatial dimensioness is zero, so the number of spin components is $n = 0$. The number of spin components is still $n = 0$, when the $z$-axis DREP jumps sequentially a limited number of $k_\lambda (= k_\alpha + k_\beta + k_\gamma + k_\delta)$ times around the M$^+$-P. In the theory of $n = 0$, the only allowed graph is that the $k_\lambda$ jumps of the DREP of HSM $a_0$ must form a closed loop when the M$^+$-P completes the jump of the closed path composed of $k_\lambda$ fixed points on the four consecutive diagonals on the 2$\Delta$d micro-cubic lattice. From this, the HSM IE spins are derived. Therefore, the $k_\lambda$ $\vartheta_\lambda$-operation elements on the four interfaces forming a closed loop surrounding $a_0$ constitute a "sub-soft matrix" in $\mathcal{A}$, i.e. the $V_0$ cluster; and when jumping nine $k_\lambda$ steps surrounding $a_0$, a $V_8$-soft matrix $\mathcal{A}$ centered on $a_0$ is formed. It can be seen in Fig. 2 that the emergence of DREP does not change the position of the two HSMs (two M$^+$-Ps) in the random system, but it can be accompanied by two anti-parallel identical spins. DREP is also a dynamic instantaneous Ising model state that physicists have been looking for [44]. DREP is the smallest ordered space-time unit of dynamic instantaneous positive and negative charge balance embedded in random thermal vibration, and "the embedded $\vartheta_\lambda$-operating system" is represented by $n = 0$.

**Interface excitation of monoatomic metallic glass.** In monatomic metallic glass [22, 23], each HSM has two dynamic concentric tetrahedrons. The four vertices of the two largest regular triangles with a common edge in a cubic lattice are equivalent to the four vertices of a regular tetrahedron with six sides. The center of the small tetrahedron parallel to the largest tetrahedron is also the center of the cubic lattice of each metal atom [Fig. 3(a)]. The $\theta_\lambda$-state of HSM in metallic glass is three points: the vibration equilibrium center, M$^+$-P, and the tangent point of electronic orbit on the interface, are all on the same $\theta_\lambda$-angle-line. When the two SASC M$^+$-Ps synchronously deflect the $\varphi_\lambda$ angle and jump from the $\theta_\lambda$ angle-line to the $\theta_{\lambda+1}$ angle-line, the $\lambda$th excited state of the CEP at the two tangent points $p$ and $p^*$ at the $\beta$ interface between $a_0$ and $c_0$ is a second-order delta magnetic moment, in which the two $(2\pi + \varphi_\lambda)$ unclosed orbits of the two electrons surround their respective M$^+$-P in Fig. 3(b). The $\theta_\lambda$-GT of monatomic metallic glass corresponds to the state that the M$^+$-P locates at the vibration center on the $\lambda$-angle line.

**Interface excitation of polymer glass.** Refer to the mechanism of IE in metallic glass, the form of IE in general polymer materials was discovered. In the solid-liquid transition of polymer materials, there are three concentric-synchronous-dynamic regular tetrahedrons per HSM. In polymer materials such as polyethylene, two hydrogen (H) atoms attached to the central (carbon, C) atom (Fig. 4). C and H in this structure may represent any other atom or atom of a linking group. The M$^+$-P of each HSM selects the same four vertices for cyclic-jump from the eight vertices of the HSM microcubic lattice with 2$\Delta$d sides. The M$^+$-P does not rotate in the $z$-axis during its four-step diagonal walking around its own $z$-axis. However, the four-step diagonal walking can allow two H$^+$-Ps to cooperate to walk a smaller square closed loop, thus the $k_\lambda$ DREPs achieve a maximum square IE closed loop on the $x-y$ projection plane. Here, the hard repulsion between $a_0$ and $c_0$ is due to the $k_\beta$ DREPs formed tangentially to the $\beta$-interface by the $k_\beta$ SASC unclosed orbits. Therefore, the charge electron repulsion energy of the $k_\lambda$ DREPs surrounding each HSM in the $z$-space provides a new Hamiltonian in the system for HSM clustering and walking.

**HSM walking mode**. Whether it is metallic glass or polymer melt, the movement of HSM along the $x$- or $y$- or $z$-axis in 3D space depends on the choice of the four-step diagonal cycle path of the M$^+$-P of HSM on its 2$\Delta$d cubic lattice [Figs 3(e) and 4(d)]. During the glass transition [Fig. (4)], the $M^+$-Ps of all HSMs in a soft matrix spontaneously choose the same four-step diagonal circulation path on the $2\Delta d$ cubic lattice as same as the soft matrix center HSM. As the temperature increases, the $M^+$-P of each HSM has more than one (four-step diagonal) closed paths on its $2\Delta d$ lattice. In the molten state, according to the relaxation time spectrum, the $M^+$-P of the central HSM of each local area circulates nine times in sequence along each of the three closed loop paths in Fig. 4d on its $2\Delta d$ micro-cubic lattice, thereby generating soft matrices in three spatial directions. When these soft matrices disappear, the HSMs in the $x$-, $y$- and $z$-soft matrix walk $n_x$, $n_y$, and $n_z$ steps along $x$-, $y$- and $z$-axes, respectively (A.1, Fig. 5).

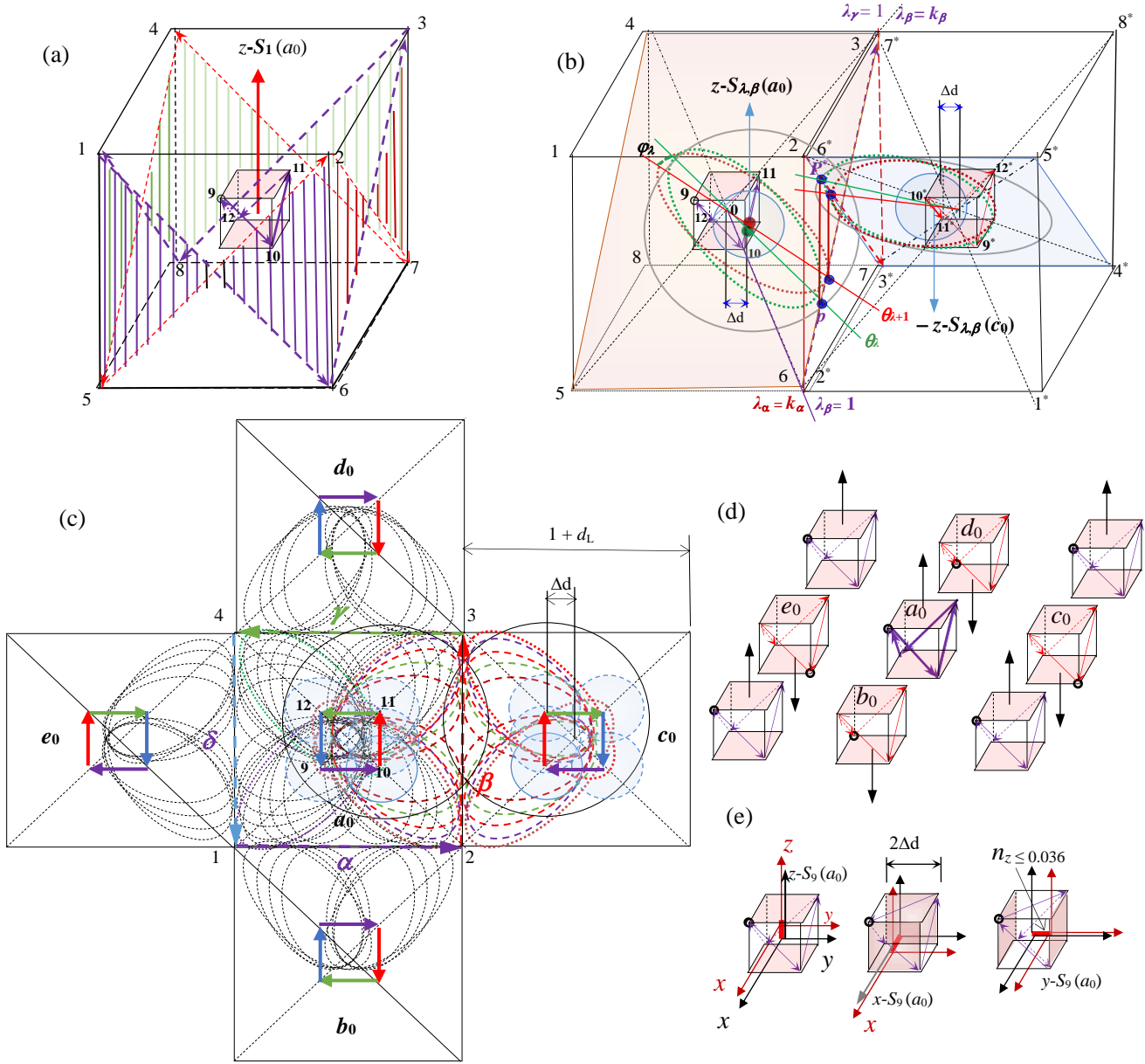

FIG. 3. Interface excitation in metallic glass. (a) 1-6-3-8-1 and 9-10-11-12-9 are two synchronously generated regular tetrahedrons. Positive charge- center particle (M$^+$-P), also HSM centroid, makes the parallel jump-transport (PJT) along the four sides 9-10-11-12-9 in the 6 sides of the tetrahedron. The diagonal planes of the four PJTs are, respectively: 9→10 at the 1-6-7-4 plane, 10→11 at the 6-3-4-5 plane, 11→12 at the 3-8-5-2 plane and 12→9 at the 8-1-2-7 plane. (b) $a_0$-$c_0$ coupling diagram; all SASC electron orbits are only located in the two orthogonal diagonal planes. The two $\theta_\lambda$-SASC electron orbits are tangential to the two-atomic interface, and the two tangent points create the $z$-axis $\lambda$th-DREP. At the next moment, the two $\theta_{\lambda+1}$-SASC electron orbits are tangential to the interface to produce the $z$-axis ($\lambda$+1)th-DREP. (c) The 2D projection of Fig. 2(b) assumes that the start and end points of the M$^+$-P cycle in Fig. 2(a) are at point 9. (d) Each M$^+$-P of the four adjacent atoms of $a_0$ jumps in exactly the same way as the center atom $a_0$ except for the direction and starting point (phase) of the cycle, marked with small black circles. The four adjacent HSMs $b_0$, $c_0$, $d_0$ and $e_0$ have the same spin as HSM $a_0$, but the opposite direction. (e) The spins that generate the $x$ and $y$ axes of the HSM are only caused by the choice of the four sides of the six sides of the micro-cubic lattice. The same cycle start and end points (small black circle) can have three loop paths to produce three zero-spin components (and three displacements in the $x$, $y$, and $z$ axes). When an axial soft matrix of $a_0$ is generated and disappeared, the new local coordinate system (red) moves $n_z$ steps relative to the original coordinate system (black).

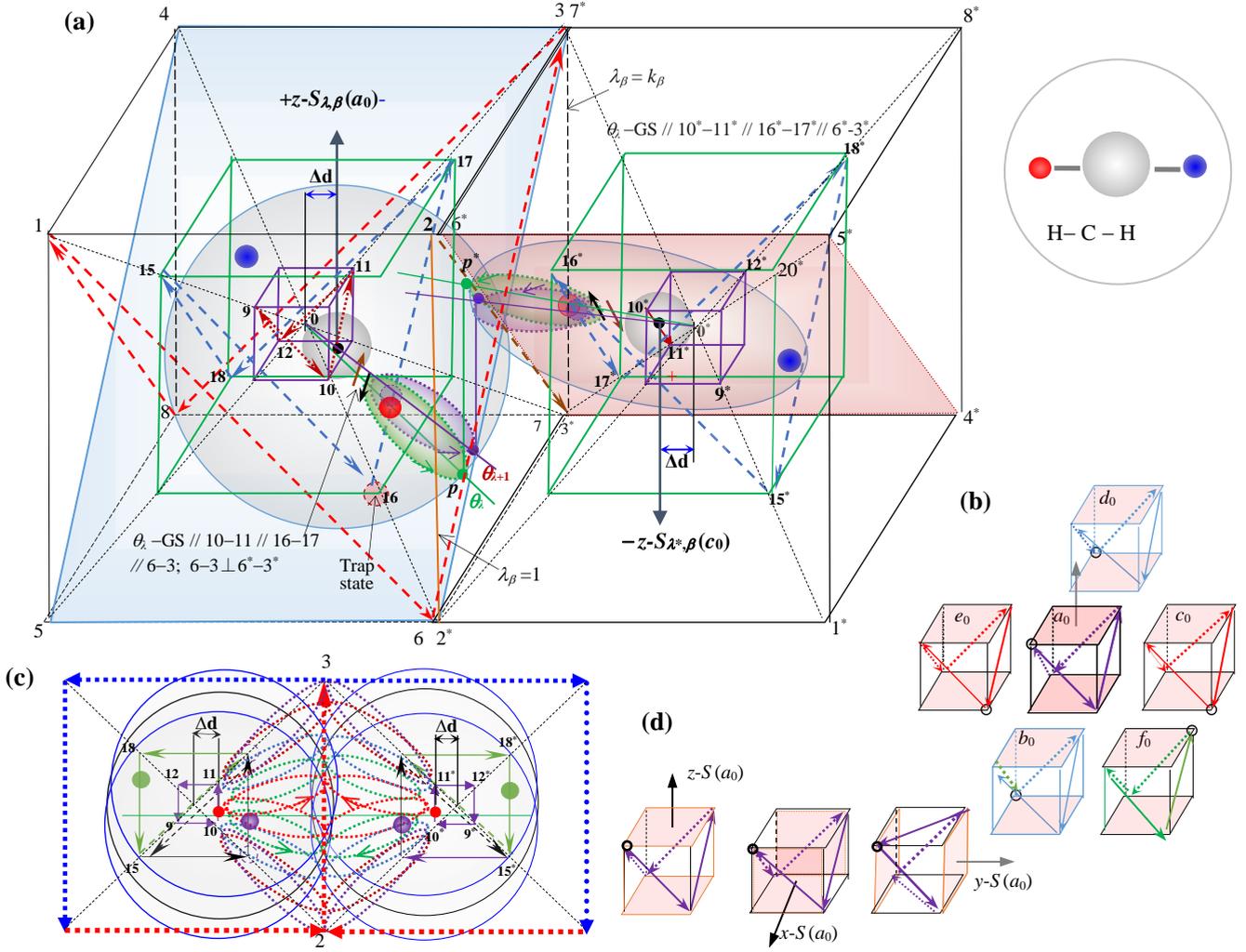

FIG. 4. Synchronous antisymmetric coupling electron orbits and the $\beta$-IE between the HSMs $a_0$ - $c_0$. (a) The green $\theta_\lambda$-line indicates that the four points: the center of HSM cubic lattice (HSCL), the instantaneous position of HSM positive charge-center particle ($M^+$-P), also the HSM centroid, the instantaneous position of hydrogen atom positive charge-center particle ($H^+$-P) and the tangential point of the orbit at the $\beta$-interface are on the same $\theta_\lambda$-angle-line. The diagonal 0-6 of the 60 degree angle at which the diagonal plane (4-1-6-7) and the diagonal plane (4-5-6-3) intersect is the trap state, where $M^+$-P and $H^+$-P take a long time (through the fluctuation of the angle) to complete the 60 degree angle conversion from $\lambda_\alpha = k_\alpha$ state perpendicular to the diagonal plane (4-1-6-7) to $\lambda_\beta = 1$ state perpendicular to the diagonal plane (4-5-6-3). And the 0–3 angle-line is also a time-consuming state in which red $H^+$-P leaves 0–3 angle-line, blue $H^+$-P reaches 0–3 angle-line due to random vibration, resulting in $\lambda_\gamma = 1$ on $\gamma$ interface. (b) The four-step jumps of the centroids of all HSMs are arranged in a regular dynamical Ising lattice in the $x$-$y$ projection plane. The $M^+$-P of the HSM in the $z$-axis spin has only four phase-path (the start and end points of the loop) choices along the four diagonal jumps, which are four $2\Delta d$ small cubes: $a_0$, $b_0$, $c_0$ and $f_0$. The black dots on the HSCL vertices indicate the phase-path. (c) Fig. 3(a) projection in the $x$-$y$ plane. The $k_\beta$ points that appear in order from 2 to 3 form an arrow of the $\beta$–IE. (d) The same cycle start point (small black circle) for the centroid ($M^+$-P) of each HSM in the melt can have three different four-diagonal-jump circulation paths, which will result in three HSM IE spin components along the $x$-, $y$-, and $z$-axes, and walking in three directions.

**Interactions and clustering bonds.** The existence of CEP in a molecular system is common sense. There are many CEPs in a system. However, if CEP appears on the interface shared by two HSMs, it is a clustering event of the two HSMs. This may also be the mode in which two electrons meet in a physical and chemical reaction. Without the theory of $n = 0$ we cannot find the mode where the two electrons meet. At best, we know that two electrons meet at the interface with the way of opposite spins and a repulsive energy. According to the strict $n = 0$ theory, except for the SODVs in Fig. 2, all coupled electron pairs will not appear at the interface of two HSMs. The meeting mode of the two coupled electrons on the $u$ interface (including trap state) is the $k_u$ non-monotonic discrete instantaneous excited states of CEP, which is completely different from the excited state mode of electrons. The excited state energy [~0.56 meV [24], see Eq. (34))] of the CEP is three orders of magnitude lower than that of electrons (~ eV).

The image of $u$-IE exported in this study is as follows, from the trap state of $\lambda_u = 1$ to the trap state of $\lambda_u = k_u$ in Fig. 4(a), the $k_u$ $z$-axis instantaneous excited states ($z$-axis DREPs) of CEP appears in turn, and $k_u$ $\varphi_\lambda$ non-closed orbit pairs are also accompanied on the $x$-$y$ projection plane [Fig. 4(c)]. Therefore, when DREP surrounds a closed loop of $a_0$, the $k_\lambda$ $\varphi_\lambda$ unclosed orbits of the four interfaces surround M$^+$-P and two H$^+$-Ps in $a_0$ and form a closed loop magnetic moment. The singularity is that the $\beta$-IE vector can occur $m$ times according to the relaxation time spectrum $\tau_m$ ($= \tau_{i+1}$); and when the IE vector appears for the $m$-th time, the $m-1$ IE closed loops surrounding $a_0$ and $c_0$ have been completed respectively, resulting in the interaction between $a_0$ and $c_0$ being the spin interaction in the dynamic Ising model, that is, HSM IE vector is a clustered bond between two adjacent HSMs.

**The soft matrix explanation of various glass transition phenomena.** The discovery CEP excited states has injected vitality into the glass model of de Gennes. Without other assumptions, the 2D soft matrix spin system can directly explain many phenomena in the glass and glass transition, including free volume [24-26] (C.4 and G); jamming [27, 28] (B.1, Fig. 6); trap [29], Fig. 4(a); energy landscape [30-33]; heterogeneity [34,35] (C.4); amorphous rigidity [36]; cluster movement zero entropy temperature (B.3, Fig. 8); random first-order transition [37] (F. Eq. 27); anomalous relaxation law of glass state (E. Eq.25); ideal glass transition [38,39] (B.3, Fig. 8); thermodynamics [40,41] (F. Fig. 26); replica symmetry (A.2), replica symmetry breaking caused by the magic number 14 (Fig. 8), and 20-fold symmetry of IE in random system (B.3, Fig. 8).

**Amorphous stiffness.** The origin of rigidity in the solid state is an old issue [42], and the molecular process of obtaining the amorphous rigidity of the liquid upon cooling is not fully understood [43]. $n = 0$ CEP provides a new perspective on the mystery of stiffness. If external stress is applied on the $z$-axis, the system will randomly excite many $z$-axis 2D soft matrices. The trajectory of each HSM centroid (M$^+$-P) in the soft matrix is a regular tetrahedron in a $z$-axis $2\Delta d$ microcubic lattice. On the $x$-$y$ projection plane, the $k_u$ transient excited states of two coupled electrons between every two adjacent HSMs forms a $u$-hard-repulsive interface between the two HSMs. And any tiny unit deformation along the $z$-axis (about $n_c \leq 0.036$) must reach the energy $k_B T_m° = k_B T_m$ that destroys the solid stiffness and the energy consumption $k_B T_g$ per unit volume.

**Space-time symmetry of random molecular systems.** $n = 0$ SODVs reveal that the space-time symmetry common to all random molecular systems will lead to various glass transition phenomena. At 9 time points in the relaxation time spectrum, the random vibration of the molecules will not change the direction of the $\mathbf{V}_\lambda$ angle-line vector of the positive-negative charge balance in HSM. Explain in detail, on the $\beta$-interface between two adjacent HSM $a_0$-$c_0$, when the $\lambda$th-DREP appears at local time $t_{0,\lambda}$, the $\lambda$th-DREP will connect two instantaneous spins with the identical spin and opposite directions, $S_{\lambda,\beta}(a_0)$ and $S_{\lambda*,\beta}(c_0)$ in Fig 4(a), this is an instantaneous Ising model state in which the $\lambda$th DREP connects the two HSM IE spins of molecular size and $\tau_0$ relaxation time. The space-time symmetry indicates that at the other 8 $t_{i,\lambda}$ ($i = 1, 2, 3 ... 8$) moments, the $\lambda$th-DREP on the $\beta$-interface must appear again. At these points in time, the $\lambda$th-DREP connects two spins, $+z$-$S_{i+1}(a_0)$ and $−z$-$S_{i+1}(c_0)$, these are also equivalent to two increasingly large instantaneous clusters $V_i(a_0)$ and $V_i(a_0)$, which can reach the nanometer scale and $\tau_8$ relaxation time. This means that the $\lambda$th-DREP at these time points plays the role of forming a set of dynamic Ising models of the cluster $V_i$ scale. These dynamic Ising models of each CEP excited state (DREP) in HSM $a_0$ can be applied to the relaxation times from $\tau_0$ to $\tau_8$ (or macro time-scale at low temperature) and from molecular size to nanoscale [$S_9$ ($a_0$) is equivalent to $V_8$ ($a_0$), Figs 6–9]. This is *the global nature of each CEP excited state*, which can explain this doubt: why not take on the configuration with the lowest energy [42]. This view is consistent with the experimental results that the enhanced dipole-dipole force from the melt leads to better glass formation [44]. The idea of a 2D soft-matrix spin system with amorphous stiffness derived from $n = 0$ CEP may be expected to be confirmed or denied in the latest 2D amorphous materials [45].

**The theoretical origin of relaxation time spectrum.** Different from other glass theories, the relaxation time spectrum in the soft matrix comes directly from all $\theta_\lambda$-$\theta_{\lambda*}$ states of the positive and negative charges within all HSMs in the system, each relaxation time $\tau_i$ is the time required to form a 2D-$V_i$-IE closed loop in a random process, ranging from the highest C–H bond vibration (higher than the Debye vibration frequency of the crystal) to the macro time scale. The minimum relaxation time for the balance of positive and negative charges in metallic glass is the unclosed orbital period of the coupled electrons in Fig. 3(b). The theoretical origin of such a wide range of relaxation times has also been the subject of theoretical scholars' attention [3].

**The renormalization in $n = 0$ theory.** The HSCL formed by the four $z$-axis CEP excited state interfaces is the excluded volume of HSM. The concept of excluded volume is at the heart of much of polymer physics. When de Gennes $n = 0$ was discovered, many scholars discussed and verified it in [2]. Unfortunately, many scholars misunderstood $n = 0$ as the exponent of the excluded volume in the Wilson theory. The consequences of this misunderstanding may have caused a generation to ignore the influence of Gennes $n = 0$ theory on glass theory. All excited states of CEPs in $z$-space strictly give all 320 [of which four IEs form $S_{9A}(a_0)$ and the other four form $S_{9B}(d_0)$ in Fig. 9] dynamic hard-repulsive interface states, which constitute the excluded volume of HSM at different relaxation times and spatial positions. These states also reveal the difference between the $n = 0$ theory and the Wilson theory, that is, in the $n = 0$ theory, the renormalization of 2D clusters $V_i$ of different scales corresponds to the $(i+1)$-th appearance of DREPs on HSCL, instead of the excluded volume exponent of zero in the Wilson theory, i.e. $S_m(a_0)$ is equivalent to $V_i(a_0)$, $m = i + 1$.

***k* vector in *n* = 0 theory.** When looking for the source of IE, we discovered the spin system of the excited states of CEPs, which may be a new version of the statistical mechanism of the glass theory predicted by Anderson [46]. Anderson once stated: "The deepest and most interesting unsolved problem in solid state is probably the theory of the nature of glass and glass transition"[47]. Quantum chemistry shows that in random thermal vibration, no matter where the instantaneous M$^+$-P position of HSM is, the electron in a hydrogen atom connected to a carbon atom is always in a stable C–H bond resonance state. However, this study shows that, according to the strict de Gennes $n = 0$ theory, an exception occurs when two adjacent HSMs are clustered, and this exception will construct a 2D soft matrix to realize the de Gennes glass model. It should be noted that the $\mathbf{V}_\lambda$-vector (positive charge angle-line**)** in this study is the *k* vector in $n = 0$ theory: the only non-zero $k^2$ term refers to the $\mathbf{V}_\lambda \cdot \mathbf{V}_{\lambda^*} = \delta_{\lambda\lambda^*}$. The term "soft matrix" in this study refers to the dynamic 2D lattice formed by all the $z$-axis excited states of CEPs.

The 2D soft matrix also conforms to the simple spin glass model of Edwards and Anderson: to incorporate the two physical ingredients of geometric frustration and quenched disorder into the lattice Hamiltonian and Ising models [48].

## III. CONCLUSIONS AND OUTLOOK

In order to prove the abnormal viscosity of the entangled polymer melts, we discovered the excited states of CEP in random systems, which also provides an opportunity for the development of glass theory. The excited states of CEPs appear in a way that the mean field HSMs have IE spins, which is also an expression of the excluded volume of HSM. HSM IE spin will provide a new way to explore chiral spin in life systems. CEP excited states is to provide a new perspective for solid state physics: in addition to the electronic excited states expressed in energy levels, there are also excited states of coupled electron pairs described by soft matrix spin systems. Each CEP has $k_u$ transient interface excited states, which appear in sequence to form the *u*-IE vector between the two HSMs. The important feature is that the *u*-IE vector can appear *m* times on the *u* interface with the relaxation time $\tau_m (=\tau_{i+1})$. The concept of instantaneous interface excited states of CEP unifies the molecular interaction, clustering, cluster movement, rigidity, and excluded volume and relaxation time spectra in the HSM model in spin glass and polymer physics.

CEP excited states give us a new understanding of the temperature $T$ describing the degree of random thermal vibration of molecules. In a system at any temperature $T$, the energy of a large number of nanoscale 2D soft matrices embedded in the random molecular system (or HSM IE spin $S_m$ under low temperature conditions) and the $kT$ disordered vibration energy are always in balance. When $T$ rises to $T + \Delta T$, a new ideal random system embedded with more ordered 2D soft matrices balanced with the random energy $\Delta kT$ will undergo an instantaneous first-order space-time phase transition [49]. The theoretical proof of the famous WLF experimental law confirms this view [50] (F. Eq. 26).

CEP excited states reveal the existence of neighborhood effects caused by "unpredictable space-time geometry" in various amorphous molecular systems, such as, polyatomic metallic glass, polymer solutions, surfactants, gel particles, proteins, etc. Each random system has its own unique space-time symmetry. That is, the neighbourhood effect can create a spatiotemporal ordered structure. The task of studying the excited state form of CEP in each system (eg, polymer solution) is still very difficult. However, we already have a paradigm of soft matrix. The 2D soft matrix of each material contains information about the molecular chemical structure of the material, including $T_k$, $T_g$ and $T_m$. This information is contained in the excited states of the CEPs surrounding each HSM, including the four trap states of the HSM. The relationship between molecular structure information and CEP excited states will be a new research topic.

The CEP excited states reveal that the Hamiltonian in the random system is the average ordered energy of the oriented excited states of all CEPs, $H = k_B T_g$. Thus, in the new perspective, the glass transition is regarded as the disorder energy ($kT_g$) of the system, which balances the ordered energy generated by all SASC positively charged pairs and all SASC electron pairs in the system.

The way in which CEP excited states appear is the confluence of both the thermodynamic and the kinetic dimensions during liquid ↔ glass transition process, and this way has always been "one of the most formidable problems in condensed matter physics"[51]. The theory of de Gennes $n = 0$ gives: the only way the system allows CEP excited states to exist is that they appear one after the other (kinetics, where each state is an instantaneously ordered Ising model state), and form one to nine DREP closed loops surrounding each HSM to form a $z$-axis 2D soft matrix (thermodynamics). The confluence of both the thermodynamic and the kinetic dimensions is the basic feature of life sciences. This means that the soft matrix concept of CEP excited states will be widely applied to soft matter systems including biomolecules.

The excited states of CEPs depend on the five-HSM-five-cluster-five-local-field model. In the five-HSM model, the physical quantities that can carry all the chemical structure information of the five HSMs are the closed-loop walking of SASC M$^+$-P pairs and the energy landscape of SASC CEPs, which can be used to interpret and prepare many new materials, for example, a new class of high-entropy alloys consisting of five equal-atom elements [52-55], and an ideal polymer/pentamer mixtures [56].

The concept of CEP excited state derived from de Gennes $n = 0$ will become part of the soft matter theory created by de Gennes. In particular, since the energy of all CEP excited states constituting the 320 interfaces of the soft matrix is the *orientation activation energy* of the CEP excited states (see the theoretical proof of the WLF equation (F. Eq. 27), therefore, the soft matrix can also be an intermediate model that must pass through when two electrons meet at the interface in physical and chemical reactions. This indicates that the glass theory established by CEP excited states will also be a theoretical tool for chemical biology, which can be applied to disordered protein interactions [57] and intracellular near-glass metabolism [58] and cytoplasmic glassy behavior [59, 60] etc.

We emphasize once again: the global nature of each CEP excited state is the core concept of glass theory. The inevitable result of this research is to correlate the potential energy landscape of each material (including biological materials) with all the excited states of the CEPs in the material. Although there are still many difficulties, this may be the future trend of theoretical and experimental research on glass and soft matter, just like dealing with the electronic behavior and electronic energy levels of materials.

## APPENDIX

## A. Introducing IE to Prove 3.4 Power Law

### 1. Polyester melt super-high-speed spinning experiment results sprouted the idea of soft matrix

Molten polyester can transform entangled random macromolecules into structurally stable fully oriented polyester fiber (the glass transition is complete) within a few milliseconds under super-high-speed spinning conditions. However, under normal spinning speeds, the resulting fibers are not fully oriented and have unstable structures, and it takes hours or even days to complete the glass transition. The super-high-speed spinning of the polyester melt links the entangled polymer melt viscosity and the glass transition in one direction in milliseconds. One possible explanation is that only the largest 2D cluster (soft matrix) can move. There is a plurality of spatially oriented soft matrices in each localized region of the melt. The molten super-high-speed spinning is oriented in the same direction on the spinning line for all soft matrices in the supercooled liquid. A normal glass transition involves the formation of a soft matrix in a certain direction in each localized region. In the super-high-speed spinning of polyester, within a few milliseconds, the temperature of the molten filament dropped from 300 degrees to room temperature and was stretched by more than 30,000%, which indicates that the formation of the soft matrix is independent of temperature and material deformation. Therefore, understanding how molecules form clusters may constitute a breakthrough in the glass problem.

Experimental data on the cooperative orientation activation energy, $\Delta E_{co} = 2035$ ($k_B T$), of an on-line measurement of polyester under super-high-speed spinning [61] supports this view. In the experiment, $1/6\ \Delta E_{co} = 339$ ($k_B T$) is the energy of the glass transition temperature $T_g$ ($\approx 339$ K$\approx 67°C$) of polyester. The coefficient 1/6 is derived from the ideal random orientation distribution of macromolecules in the melt. This may mean that the average orientation energy of a soft matrix, whether in the melt or in the glass state, should be $k_B T_g$.

### 2. Molecular clustering and replica symmetry require the concept of IE

In the five-HSM model, the central HSM $a_0$ with L− J potential $f(\sigma / q)$ has a hard-sphere $\sigma$ and a IE closed loop composed of four excited interfaces appearing in sequence and surrounding $a_0$, written as +z-$V_0$ ($a_0$)-loop (Fig. 6). This is a new dipole moment vector. In statistics, the $\mu$-$V_0$ ($a_0$)-loop in any $\mu$-direction of $a_0$ interacts with the $V_0$-loops in all different directions in 3D space to form a new L− J potential $f_0$ ($\sigma_0 / q$) with the hard-sphere $\sigma_0$ in the $a_0$ local field. The IE closed loop formed by the four interfaces surrounding $a_0$ in the five HSM gaps in the $V_0$ ($a_0$)-cluster is the IE spin of $a_0$, +z-$S_1$ ($a_0$). The IE closed loop represents the hard repulsion between two adjacent HSMs [Figs. 3(c) and 4(c)]; therefore, the $V_0$ ($a_0$)-cluster still contains only 5 $\sigma$. The four adjacent HSMs are replicated using positional fluctuation symmetry similar to [62] where the adjacent HSMs have the same orientation structure as the central HSM. After $a_0$ local time $t_0$ ($a_0$), $\mu_b$-$V_0$($b_0$), $\mu_c$-$V_0$ ($c_0$), $\mu_d$-$V_0$ ($d_0$) and $\mu_e$-$V_0$ ($e_0$) are sequentially obtained by replicating $V_0$ ($a_0$) in the $\mu_b$- and $\mu_c$-…direction [A.1, Fig. 9(a)] at local times $t_0$ ($b_0$) and $t_0$ ($c_0$)…. Similarly, in Fig. 1 at time $t_1$ ($a_0$), $\mu_b$-$V_0$ ($b_0$), $\mu_c$-$V_0$ ($c_0$), $\mu_d$-$V_0$ ($d_0$) and $\mu_e$-$V_0$ ($e_0$) are again projected in the $z$-$a_0$ field one after the other to couple with $V_0$ ($a_0$) to form a new 2D −z-$V_1$ ($a_0$)-cluster [and 2D −z-$V_1$ ($a_0$)-loop] with 17$\sigma$. Furthermore, the $\mu$-$V_1$ ($a_0$)-loop interacts with all $V_1$-loops in different directions in 3-D space to form 3D hard sphere $\sigma_1(a_0)$ with 17$\sigma$ in the $a_0$ field and a new L− J potential $f_1$ ($\sigma_1/q$), see Fig. 1. The numbers of $\sigma$ and IEs required to form +$V_2$ ($a_0$), −$V_3$ ($a_0$), +$V_4$ ($a_0$), −$V_5$ ($a_0$), +$V_6$ ($a_0$), − $V_7$ ($a_0$) and +$V_8$ ($a_0$) in the inverse cascade are directly obtained [12, 13]. It can be seen that the purpose of introducing the IE concept of HSM is to make each 2D cluster still a 2D IE–loop magnetic moment, so that it can generate a larger 2D cluster with other 2D clusters.

### 3. Fast-slow (normal-abnormal) interaction in clustering

We establish nine long-range and fast-acting reduced [potential well energy $\varepsilon_0(\tau_i)$ unitization] L–J potentials represented by $\sigma_i$:

$$f_i(\sigma_i/q_i) = 4[(\sigma_i/q_i)^{12} - (\sigma_i/q_i)^6] \quad (1)$$

Within the relaxation time $\tau_i$ of $a_0$, a short-range and slow-acting additional $\pi$-phase potential is introduced:

$$\chi_i = (\sigma_i/q_i)^6 \quad (2)$$

It can be seen from $\chi_i = 1$ that, like the "unit 1" of measuring $\Delta d$, the "unit 1" of measuring $\chi_i$ also comes from these states of all hard repelling clusters $\sigma_i$ ($i = 0, 1, 2…8$) measured by the unit DoF energy $\varepsilon_0$. The physical meaning of $\chi_i$ potential is that when $V_{i-1}(b_0)$, $V_{i-1}(c_0)$, $V_{i-1}(d_0)$ and $V_{i-1}(e_0)$ are sequentially coupled with $V_{i-1}(a_0)$ at $\alpha, \beta, \gamma$ and $\delta$ interfaces in the $x$-$y$ projection plane (Fig, 3c), the L–J potential $f_{i-1}$ suddenly superimposes an extra potential $\chi_i$ caused by the $2\pi$ closed loop surrounded by the four $V_{i-1}$ clusters (the neighborhood effect). That is, the $a_0$ field has a fast-slow (long range − short range) interaction, and the relationship between the two is:

$$f_i(\chi_i) = -4\chi_i(1 - \chi_i) \quad (3)$$

The balance of the two potential fluctuations are

$$\Delta f_i(\chi_i) = \Delta \chi_{i+1} = \Delta \chi_i (\partial f_i/\partial \chi_i) \quad (4)$$

The stability conditions [13] are:

$$|\Delta \chi_{i+1}/\Delta \chi_i| \leq 1 \quad (5)$$

### 4. Fixed points, Lindemann ratio $d_L$ and clustering constant $\Delta d$

From Eqs (1) to (5), we obtain nine fixed points of nine L–J potentials: $f_c = 1/16\,\varepsilon_0(\tau_i)$ at the nine cluster positions on the $q$-axis in Fig. 1; and $\chi_{min} = 3/8$, $\chi_{max} = 5/8$.
In the vibration equilibrium position,

$$\chi_0 = 1/2, \quad q_0 = 2^{1/6}\sigma$$

At nine fixed points:

$$q_{i,R} = q_{i+1,L} \quad (6)$$

$$\sigma_i = \sigma_1 (5/3)^{(i-1)/6}, \ i = 1, 2, 3... \quad (7)$$

$$\Delta q_i = q_{i+1,R} - q_{i+1,L} = q_{i+1,R} - q_{i,R} = (8/3)^{1/6}[(5/3)^{1/6} - 1]\sigma_i$$

$$\Delta q_i \approx 0.1046\,\sigma_i \quad (8)$$

Eq. (8) gives Lindemann ratio $d_L$ and clustering constant $\Delta d$

$$d_L = \Delta q_i / \sigma_i = (q_{iR} - q_{iL})/\sigma_i$$

$$= (q_R - q_L)/\sigma = 0.1046… \quad (9)$$

$$\Delta d = (q_{i\,R} - q_{i\,0})/\sigma_i = (q_R - q_0)/\sigma \approx 0.055 \quad (10)$$

The clustering constant is the distance $\Delta d$ from the right fixed point to the center of vibration balance. Eq. 10 represents a dynamic microcubic lattice with $2\Delta d$ sides inside the mean field HSM, which is composed of a jumping closed-loop trajectory of the centroid (positive charge center particle, $M^+$-P) of each HSM.

### 5. The step size of the soft matrix walk

To describe the step size of the soft matrix walk in either a glassy or liquid state, it is necessary to find the probability $n_z$ that $a_0$ occupies the central cavity space (the "vacancy" volume (Eq. 9) required for HSM walking) of its soft matrix [12]. As de Gennes said [5], the "vacancy" is not empty, but is full of low density matrix. The IE closed loop appears 9 times on the four interfaces of HSCL $a_0$, causing $a_0$ to jump out of the potential well and leave a "vacancy", labeled $O(a_0)$. The volume of $O(a_0)$ is measured with energy $\varepsilon_0 = 1$. $n_z$ is also the probability that $a_0$ as a circulation singularity (destroying solid lattice) occupies the largest IE closed loop $V_8(a_0)$-loop. This step size $n_z$ can essentially be regarded as the update distance of the local coordinate system during the soft matrix generation and reconstruction.

In Fig. 5(a), Edwards' production operator $\hat{p}_+$ and disappearance operator $\hat{p}_-$ act on $a_j$; that is, the probability of generating a $+z$-$V_8(a_j)$ soft matrix from time $t_8(a_j)$ to $t_8(a_j) + \tau_8$ is $\hat{p}_+$.

$$\hat{p}_+ = (1/N)^{T_g/\varepsilon_0} \quad (11)$$

Where $1/N$ comes from the Edwards $z$-component tube model modified by the author (D.1, Fig. 5), which means that all the different spatiotemporal interactions between the reptation chain in the centerline of the $z$-component tube and all surrounding chain HSMs are represented by $200N$ $z$-axis HSCLs, which have equal chances of generating $N$ tube cross-section soft matrices sequentially. $T_g/\varepsilon_0$ indicates the number of equivalent particles that the $V_8$-loop has when using $\varepsilon_0(\tau_8)$ to measure the energy ($k_B T_g$) of the soft matrix $V_8$-loop. Note that in a soft matrix composed of clusters in reverse cascade, the average energy ($k_B T_g$) of the soft matrix is concentrated in the $V_8$-loop.

The possibility of eliminating the soft matrix of $+z$-$V_8(a_j)$ from time $t_9$ to $t_9 + \tau_8$ is $\hat{p}_-$.

$$\hat{p}_- = (n_z)^{T_m/\varepsilon_0} \quad (12)$$

Where $T_m/\varepsilon_0$ represents the equivalent particle number for updating the energy of the $a_j$ soft matrix (Fig. 5). $\varepsilon_0 = \varepsilon_0(\tau_8)$ is the potential well energy of the $V_8$ cluster. This a material parameter that does not change with system temperature. Thus

$$\hat{p}_+ = \hat{p}_- = \boldsymbol{p}$$

$$(1/N)^{T_g/\varepsilon_0} = (n_z)^{T_m/\varepsilon_0}$$

$$n_z = (1/N)^{T_g/T_m} \quad (13)$$

For flexible macromolecules, $T_g/T_m = 5/8$, Eq. (20), and the critical number average molecular weight $N_c \approx 200$, from

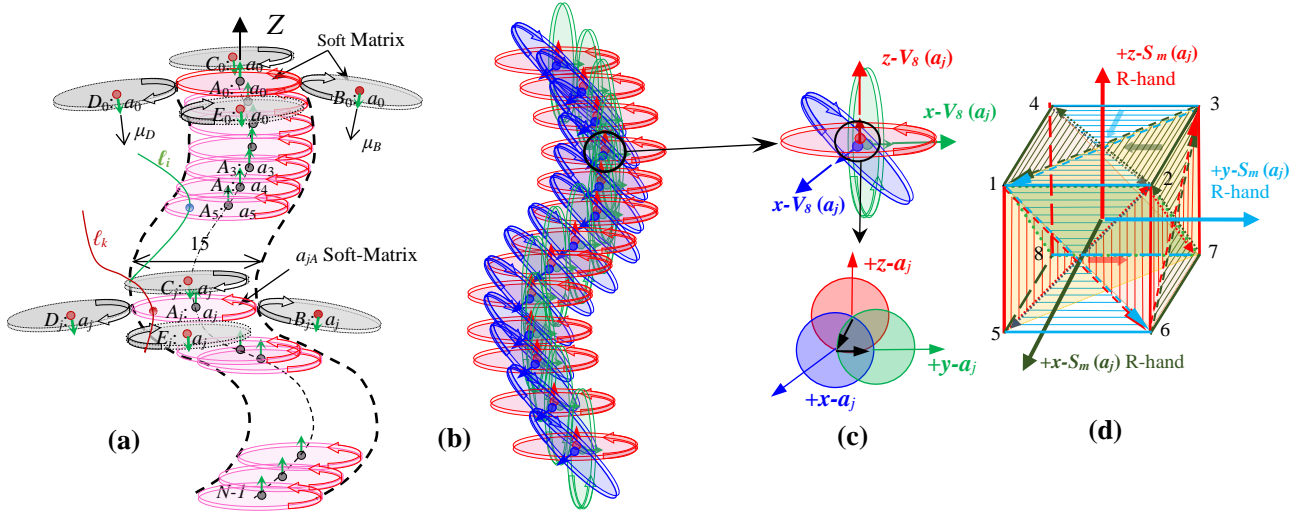

FIG. 5. **Soft matrix jump mode.** (a) Modified Edwards $z$-component tube (tube length is equal to the chain length $N \geq N_c \approx 200$, the reptation chain is located on the centreline of the tube). Edwards' production operator $\hat{p}_+$ acting on the 5 centre HSMs $a_j$ in 5 domains $A_j$, $B_j$, $C_j$, $D_j$ and $E_j$, respectively, during five time interval overlaps, i.e., $(t_{jA}, t_{jA}+\tau_8)$, $(t_{jB}, t_{jB}+\tau_8)$, $(t_{jC}, t_{jC}+\tau_8)$…, $j \in 0,1, 2... N-1$; at $t_{jA}+\tau_8$ moment, the generated $+z$-$V_8(a_{jA})$ soft matrix is the tube cross-section in the $A_j$ domain. At time $t_{jE}+\tau_8$, four soft matrices $\mu_B$-$V_8(a_{jB})$, $\mu_C$-$V_8(a_{jC})$, $\mu_D$-$V_8(a_{jD})$ and $\mu_E$-$V_8(a_{jE})$ appear. Their cooperation is equivalent to $-z$-axis soft matrix, or equivalent to the disappearance operator $\hat{p}_-$ acting on $+z$-$V_8(a_{jA})$ soft matrix. Note: the various interactions between all two adjacent chains in $z$-space (including chains like $\ell_j$ and $\ell_k$) have been represented by 200 excluded volumes (200 $z$-axis HSCLs) of different spatiotemporal HSMs in the soft matrix. (b) In the molten and supercooled liquid, since HSM $a_j$ circulates in sequence according to the three paths in Fig. 3(e) or Fig. 4(d) during $\tau_i$, when the soft matrix $z$-$V_8(a_j)$ moves $n_z$ step along the $z$-axis from time $t_9$ to $t_9 + \tau_8$, the soft matrix $x$-$V_8(a_j)$ and the soft matrix $y$-$V_8(a_j)$ will also move $n_z$ steps along the $x$-axis and $y$-axis respectively. (c) For all $j$ ($j = 0, 1, 2…N-1$), the three soft matrices of $a_j$ (eg, $+x$-$V_8$, $+y$-$V_8$ and $+z$-$V_8$) jump $n_x$, $n_y$ and $n_z$ steps in sequence (arrows) along the three axes (in general, $n_x = n_y = n_z$), and this sequence is the same for all HSMs in the chain (form three independent-entangled walking spin waves). (d) Graph of the excluded volume of HSM in the melt. HSM IE spin is divided into left and right hands. The HSCL in the melt has three axial (left-hand or right-hand) IE spins

Eq. (13), we get $n_z \leq 0.036$.

"Quantized step" $n_z$ is equal to "quantized step energy" and also equal to "walking DoF number", and three different physical quantities share the same dimensionless number $n_z$ as the smallest unit describing the walking of molecular clusters in the system.

### 6. Theoretical proof of the 3.4 power law

More than 60 years ago, physicists and chemists discovered that when the number of chain-molecules is greater than 200, there is a 3.4 power-law relationship between the viscosity and the number average molecular weight of the flexible polymer melts. The best result so far is the de Gennes reptation model, which gives a 3.0 power law relationship. Since chain length variable $N$ is a large number, the 3.4 power law is very sensitive for any modified theory of de Gennes reptation model. If taking the range of $N$ as 200 ~ 2000 and the error tolerance between theoretical value and experimental result of viscosity is less than ±30%, then a good theory should be able to give the theoretical exponential value range as 3.4 ± 0.05 for flexible polymers. At the same time, the theory should give a general exponential expression that should be consistent with the experimental results for non-flexible polymers. This is also a fine way to check up the molecular movement theory.

By Eq. (13), for a chain of length $N$, the number of DoF ($N_z^*$) required to walk $n_z$ steps along the $z$-direction in the Edwards $z$-component tube is

$$N_z^* = N^{(1 - T_g / T_m)}.$$

In the reptation model, de Gennes derives the relationship $\eta \sim N^3$ under the premise that the chain $N$ is a completely free diffusion chain, but this is not the case for chain $N$. The correcting scheme is that in the reptation model, the reference chain of length $N$ is replaced by a completely free diffusion chain of length $N^*$. Using the relationship $\eta \sim N^3$, we have

$$\eta \sim (N^*)^3 = N^{9(1-T_g/T_m)} \qquad (15)$$

Table 1 shows that the theoretical values are in good agreement with the experimental results, indicating that the prediction concept of "interface excitation" may be highly correct. In particular, the experimental temperature $T$ is always higher than $T_m$, and the expression contains $T_g$ and $T_m$ independent of the experimental temperature, indicating that the movement of molecular clusters is related to the two energies $kT_g°$ and $kT_m°$ of the soft matrix.

# Table 1. Experimental data and theoretical values of viscosity exponent

| Sample | $T_g$ K | $T_m$ K | Experimental value | Theoretical value |
|---|---|---|---|---|
| Flexible chain | $kT_g = 20/3\,\varepsilon_0$ | $k_B T_M = 20/3\,\varepsilon_0 + 4\varepsilon_0$ | 3.4 | 3.375 |
| Non-flexible chain Polypropylene | $-10°$ = 263 K | 449 K | 3.73 | 3.72 |
| Non-flexible chain trans-1,4-polyisoprene | $-65.8°C$ = 207.35 K | $6°C$ = 339.15K | 3.5 | 3.5 |
| Non-flexible chain Polybutadiene (I) | $106°C$ =167.15 K | $6\ °C$ = 267.15K | 3.4 | 3.4 |
| Non-flexible chain Polybutadiene (II) | $-92°C$ =181,15 K | $-6\ °C$ = 267.15K | 3.0 | 3.0 |

## B: Jamming and Percolation Transitions and Cluster Movement Zero Entropy Temperature in Clustering

### 1. Jamming particles in HSM clustering

An IE contacts two HSMs; thus, on the interface of $V_i\,(a_0)$ ($i =1, 2…7$), each IE also has an HSM on the outside of the interface, which is called the edge particle of $V_i(a_0)$ (Fig. 6). The number of edge particles of $V_i\,(a_0)$ with $i =1, 2…7$ are 12, 20, 28, 36, 44, 52, and 60 respectively; that is, $V_{i+1}$ has 8 (4 pairs) edge particles more than $V_i$, all of which are jamming particles. In the newly formed $V_i$, two adjacent edge particles occupy the same spatial position to satisfy the lowest energy condition of two identical spins in two opposite directions of $n = 0$, and they retain only one edge particle in the formation of $V_{i+1}$.

However, the situation with $V_8$ changes. The number of edge particles of $V_8$ is equal to that of $V_7$. The eight edge particles of the four $V_7$ that form one $V_8$ will cause a jamming transition in the glass transition. When $V_7\,(a_0)$ appears, the spontaneous connection of $V_7–V_6$ occurs in each local area in space (Fig. 8). In the $a_0$ area (field) that has been connected by $V_7–V_6$, if eight edge particles are squeezed again, it is actually equivalent to adding eight IEs with $\tau_8$ relaxation time to the system, four of which form $S_9(a_0)$, and the other eight form $S_9(a_j)$ in Fig. 8. Thus,

$$8\,\Delta_{IE}(\tau_i) = \varepsilon_0(\tau_i) \tag{16}$$

Where $\Delta_{IE}(\tau_i)$ is the average energy of each IE with relaxation time $\tau_i$ in the $V_i$-loop, and $\varepsilon_0(\tau_i)$ is the potential well energy of the $V_i$-cluster.

### 2. Mosaic structure of positive-negative charges in cluster enlargement

The concept of HSCL is to describe the excluded volume of HSM in the interaction between HSMs, which is also the IE spin of HSM. In the 2D projection plane, the growing 2D clusters will reveal the mosaic geometry of HSCLs (Fig. 7). The directions of the two $M^+$-Ps on the two sides of each IE are opposite. Since all $z$-axis IEs appear sequentially, the time-space characteristics of the four IEs surrounding each HSCL are different. In the increasing $V_i$ cluster, the positively charged central particle ($M^+$-P) in each HSCL is surrounded by the negatively charged IE-loop-flow, but the number of $M^+$-Ps along the $+z$-axis is different from the number along the $–z$-axis, This results in a mosaic structure of positive and negative charges in the clusters, as shown in Figs. 7–10. The wonder of the mosaic structure in the growing cluster is that it directly solves the geometric frustration problem in Fig. 9.

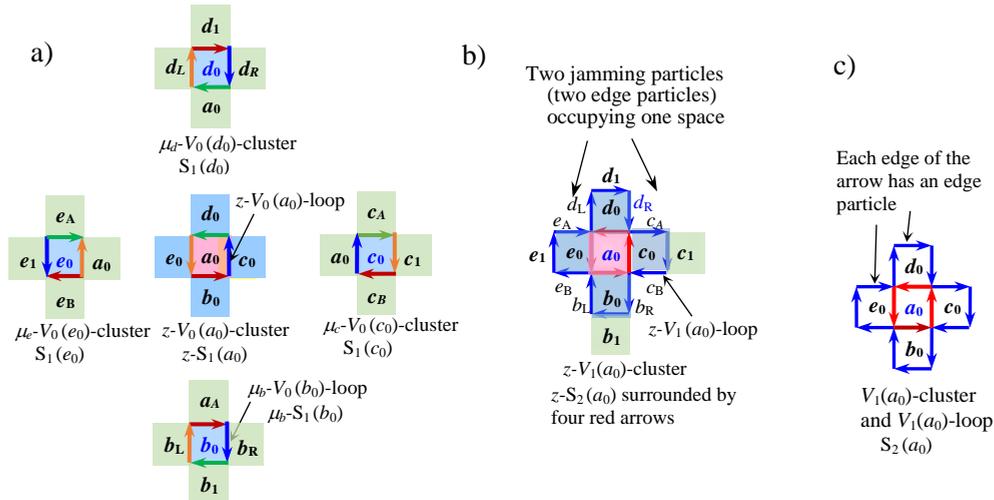

FIG. 6. $V_1$-cluster and $V_1$-loop and HSM IE spin $S_2$ in HSM clustering. (a) Four adjacent HSMs of $a_0$ ($u_0 \in b_0, c_0, d_0$, and $e_0$) generate $\mu_u$-$V_0\,(u_0)$-clusters and $\mu_u$-$V_0\,(u_0)$-loop and $\mu_u$-$S_1\,(u_0)$ in the $\mu_u$-direction of their respective HSMs. (b) Four adjacent HSMs are sequentially projected onto the $z$-axis and coupled and clustered with $a_0$ to generate a $z$-$V_1\,(a_0)$-cluster with 8 jamming particles. The 12 successive blue arrows indicate the $z$-$V_1\,(a_0)$-loop and the center four successive red arrows indicate $z$-$S_2\,(a_0)$. (c) Concise representation of the $V_1\,(a_0)$-cluster and the $V_1\,(a_0)$-loop.

frustration problem (C.1, Fig. 9).

### 3. Percolation transition in clustering and cluster movement zero entropy temperature

The introduction of IE provides a new perspective for understanding the "$T_2$ ideal thermodynamic transition or $T_K$ zero-entropy transition" below the glass transition. The key here is that the distance between the two HSCLs excited in the same orientation and the same first IE (phase) in two adjacent local areas is the *magic number* 14, which is the "mean-square end distance" of an undisturbed ideal chain composed of 196 different time-space HSCLs in the soft matrix (excluding the five HSCLs in the center, Fig. 8). As the cluster size increases, the connection between the $V_7$-IE-loop and the $V_6$-IE-loop will connect the two local areas. On the connected field of $V_7$–$V_6$, the ordered spatiotemporal dynamic structure composed of IE-loops is embedded in the gap among HSMs, which is equivalent to embedding in the thermal random vibration of $k_B T_2$ (or $k_B T_K$). $T_2 = T_g - 51.6K$ in WLF equation (26); and $T_k$ is the temperature at which the entropy change caused by the occurrence of the soft matrix is zero (no movable soft matrix). At this time, the ordered IE-loop energy of $V_7$–$V_6$ is equal to the disordered energy of thermal random vibration $k_B T_K$.

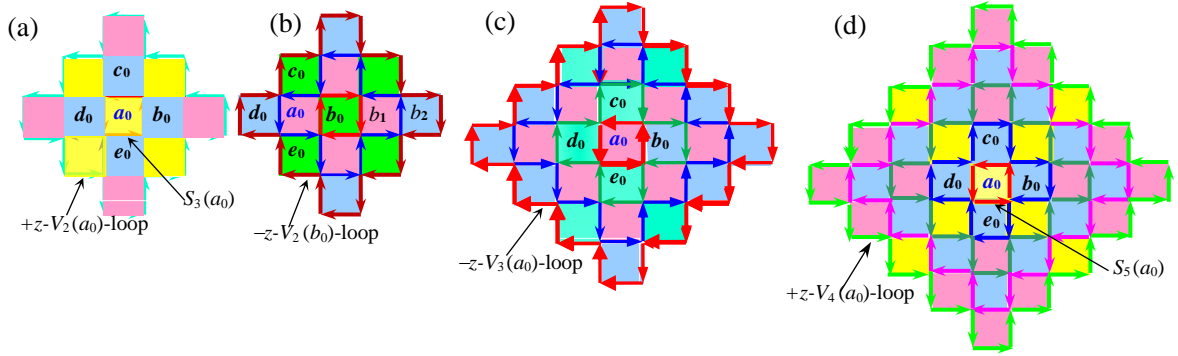

FIG. 7. Mosaic structure of positive-negative charges in cluster enlargement. (a) All negatively charged IE-loops (arrows) are located at the bottom of the L–J potential well, and all M$^+$-Ps (red, green and yellow) do not contribute to cluster movement except for 5 mosaic $+z$-direction M$^+$-P (yellow) in $+z$-$V_2$($a_0$)-cluster. (b) 5 mosaic $-z$-direction M$^+$-P (green) contributing to cluster movement in $-z$-$V_2$($b_0$)-cluster. (c) 7 mosaic $-z$-direction M$^+$-P (lake green) contributing to cluster movement in $+z$-$V_3$($a_0$)-cluster. (d) 9 mosaic $+z$-direction M$^+$-P (yellow) contributing to cluster movement in $+z$-$V_4$($a_0$)-cluster.

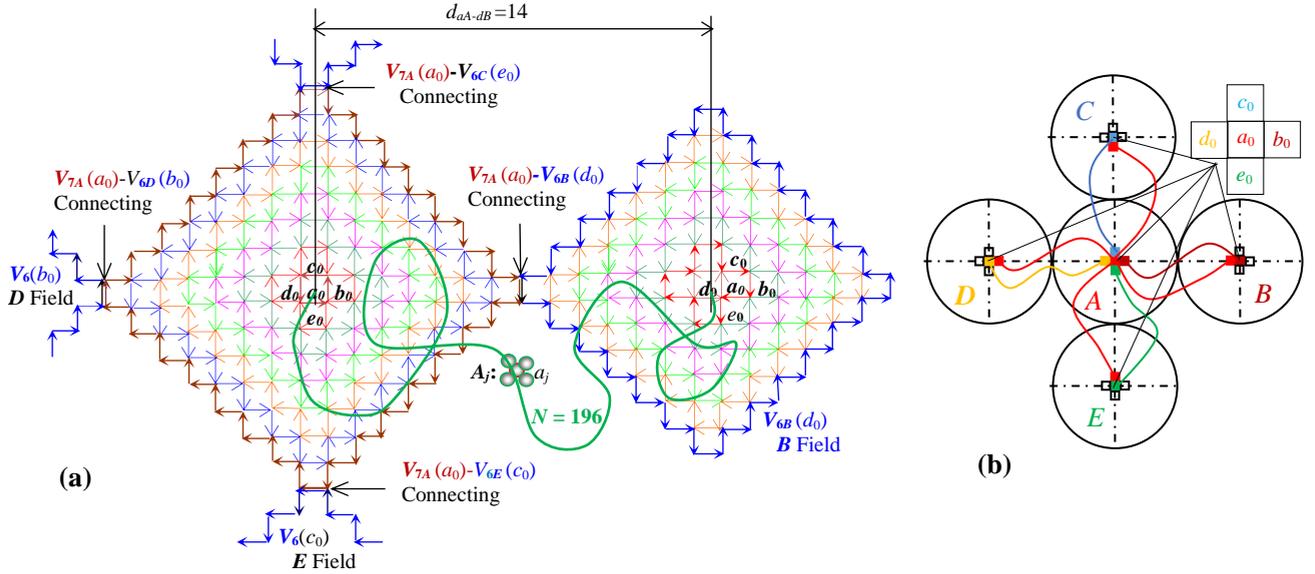

FIG. 8. Percolation transition and the 20-fold symmetry of IE in five-HSM/five-local-field. (a) Center $A$ field $V_{7A}(a_0)$-IE-loop and adjacent $B$ field $V_{6B}(d_0)$-IE-loop connection image. When $V_{7A}(a_0)$ of $a_{0A}$ appears, the $V_{7A}(a_0)$-IE-loop formed by 60 IE arrows is connected to the $V_{6B}(d_0)$-loop formed by 52 IE arrows in $V_{6B}(d_0)$. Two of the IE arrows merge into one arrow at the connected interface and reduce the two edge particles (Fig. 6) between the two clusters $V_{7A}(a_0)$ and $V_{6B}(d_0)$, thereby reducing the lattice Hamiltonian; a $V_7$–$V_6$ percolation transition occurs at temperature $T_k$, where the molecular mobility entropy becomes zero (no movable soft matrix). $N_c \approx 200 = 196 + 4$; the number 4 is the 4 adjacent HSMs of HSM $a_j$ forming $S_m(a_j)$ in an undisturbed ideal chain $N$ (green), $m < 7$. $S_m(a_j)$ also represents an HSCL state where $a_j$ interacts with surrounding HSMs and is one of the 196 HSCL states in a soft matrix (except for the five HSCLs in the center). Regardless of whether the two regions are connected by the chain $N$ or not, its two terminal HSMs $a_{0A}$ ($a_0$ in field $A$) and $d_{0B}$ ($d_0$ in field $B$) always satisfy the condition of self-avoidance random walking chain of $n = 0$: $S_{m,\alpha}(a_{0A}) \cdot S_{m-1,\beta}(d_{0B}) = \delta_{\alpha\beta}$, $m \leq 9$ in percolation transition. $196 = (d_{aA-dB})^2$. $d_{aA-dB}$ is the mean square end distance of random chain. (b) The natural connection of the fluctuation directions of all HSCLs in five-HSM-five-local field. Each color line indicates that the first IE position (phase) of the four IEs in the HSM of all soft matrix centers on the line is the same. Thus, each of the four IEs of $z$-($a_0$) in $A_0$ field is associated with a $z$-component chain (four red lines) with the same phase and the end positions of the four chains are $d_0$ in the $B_0$ field, $e_0$ in the $C_0$ field, $b_0$ in the $D_0$ field, and $c_0$ in the $E_0$ field.

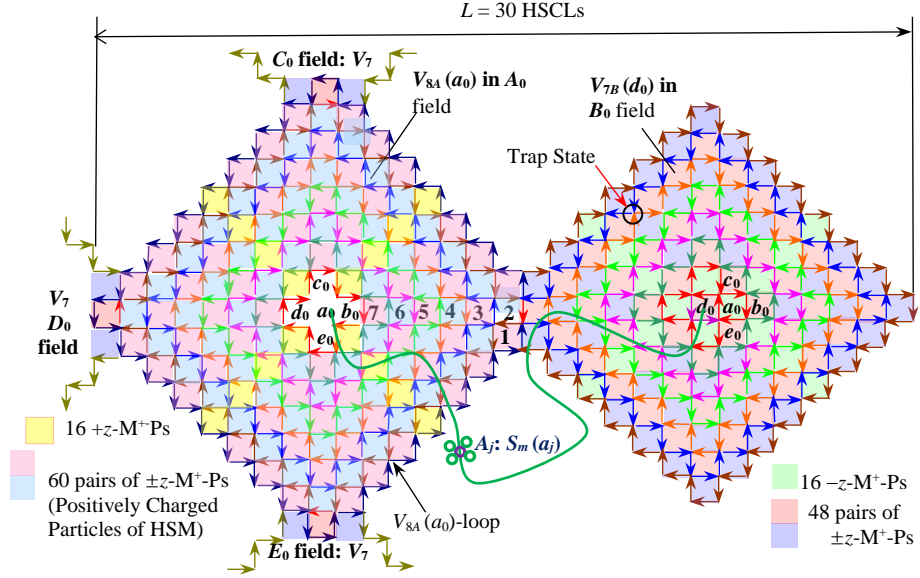

FIG. 9. Geometric frustration in soft matrix. At fixed point $q_{8,L}$ in Fig. 1, four clusters $V_{7A}(b_0)$, $V_{7A}(c_0)$, $V_{7A}(d_0)$ and $V_{7A}(e_0)$ centered on $b_0$, $c_0$, $d_0$ and $e_0$ in the $A_0$ field is sequentially coupled with $V_{7A}(a_0)$, and the largest cluster $V_{8A}(a_0)$ appears. Its average energy of 60 IEs (60 arrows) with a relaxation time of $\tau_8$ is Hamiltonian. At point $q_{8,R}$ in Fig. 1, when four clusters $V_{8A}(b_0)$, $V_{8A}(c_0)$, $V_{8A}(d_0)$ and $V_{8A}(e_0)$ appear sequentially, $V_{8A}(a_0)$ disappears, and the four sides (four arrows) of the IE-spin of $a_0$ disappear and five vacancy volumes appear. In order to reflect the contribution of the adjacent fields in the 60 IE-loop arrow in the $A_0$ field, 4 of the 320 IE states in the $a_0$ soft matrix constitute the $S_9$ of an adjacent field soft matrix, for example, it is changing from $S_{8B}(d_0)$ to $S_{9B}(d_0)$ in $B_0$ field in the glass transition..

IE spin reveals the symmetry of the first IE position in the random system. Because all local areas are dynamically connected at $T_g$, the central field in the 5-local-field model is relative (for example, the B field in Fig. 8 can also be a central field). This means that IE has "20-fold symmetry" (20 = 4 interfaces × 5 fields) in a random system.

### C. Geometric Frustration in Glass Theory

#### 1. Geometric frustration image

Geometric frustration plays a vital role in glass theory. Models that do not consider geometric frustration will not be accepted. De Gennes's evaluation of the existing mode coupling theory is that "a possible weakness of the mode / mode coupling method is ignorance of the effects of geometric frustration" [5, 6].

In the soft matrix diagram of Fig. 9, all M+-Ps and IE closed-loops (magnetic moments or IE spin $S_i$) are in the ± $z$-direction (dynamic-Ising model). When $S_9(a_0)$ in $A$ field takes the +$z$-axial direction, the extra 16 +$z$-axial M+-Ps in the $V_{8A}(a_0)$-soft-matrix are balanced with the extra 16 -$z$-axial M+-Ps in the $V_{7B}(d_0)$-, $V_{7C}(e_0)$-, $V_{7D}(b_0)$- and $V_{7E}(c_0)$-soft-matrix in the four adjacent fields $B$, $C$, $D$ and $E$ (Fig. S5), respectively, also balanced with the extra 16 -$z$-axial M+-Ps in the $V_{7A}(b_0)$-, $V_{7A}(c_0)$-, $V_{7A}(d_0)$ - and $V_{7A}(e_0)$-soft matrix in the $A_0$ field. In either case, the center position of each coupled electron pair (or each IE) (always located at the lowest energy position at the bottom of the nine L– J potential wells (the vibration equilibrium position at $z = 0$). When moving from the bottom of the potential well to the clustering fixed point of the L– J potential along the $q$-axis, each of the 16 $z$-axis M+-Ps obtains 1/16 normalized energy in the normalized potential well (Fig. 1, $\Delta\varepsilon = 1/16\varepsilon_0$). Therefore, these 16 +$z$-axis M+-Ps provide the soft matrix of the central HSM $a_0$ with DoF energy $\varepsilon_0$ moving along the +$z$-axis. The $n = 0$ SODV shows that the essence of geometric frustration is that HSM has an exclusion volume composed of CEP excited states.

#### 2. CEP excited states create new two-level systems

At temperature $T_2$ (or $T_k$), Fig. 8 shows the percolation transition of dynamic connectivity of clusters $V_7$–$V_6$ throughout the system. However, in the $a_0$ region that has been connected by $V_7$–$V_6$, by adding (press-in) 8 edge particles to form $V_8$–$V_7$ flow-percolation transition, it is equivalent to adding an extra $f_8$ potential well energy $\varepsilon_0(\tau_8)$ in the dynamically connected region of $V_7$–$V_6$. From Eq. (16)

$$\varepsilon_0(\tau_8) = 8\Delta_{IE}(\tau_8) \qquad (17)$$

The energy $8\Delta_{IE}(\tau_8)$ of the 8 jamming HSMs squeezes $a_0$ in the $z$-direction, forming a vacancy volume in the center of the soft matrix (abnormal thermal expansion), Fig. 9. In the glassy state, the thermal fluctuation energy of $T_2 + \Delta T$ causes one or a few separate (discontinuous) $V_8$ soft matrices in the system, which is the initial manifestation of the $V_8$-soft matrix, that is, soft matrix is the order parameter of glass transition order.

When the system temperature rises from $T_k$ to $T_g$, the "vacant $\varepsilon_0(\tau_8)$" in $V_7(a_0)$ of all local areas in the space will be filled in by $\varepsilon_0(\tau_8)$ and evolve into $V_8$ one by one until they are all filled. As the temperature continues to rise, $\varepsilon_0(\tau_8)$ will continue to be filled into $V_7(b_0)$, $V_7(c_0)$, $V_7(d_0)$ and $V_7(b_0)$, and melt transition occurs when all are filled. That is, $kT_m = kT_g + 4\varepsilon_0(\tau_8)$.

### 3. Hamiltonian in glass transition

In glass transition, the Hamiltonian $H$ is the average repulsive energy of all $z$-axially coupled CEP excited states newly emerging at the HSM interfaces in the system, that is, the average energy of the nine IE-closed-loops that make up the soft matrix. Since the inverse cascade does not dissipate energy, the IE energy in the soft matrix is concentrated in 60 IEs (60 arrows) in the $V_{8A}(a_0)$-loop in Fig. 9. These 60 arrows are not monotonic, among them, the 4 IE arrows in opposite directions represent the contribution of the 4 neighboring $V_7$-IE-loops in the opposite direction that make the $V_{8A}(a_0)$--loop disappear. Some (denoted as $L_{cas}$) IEs in the 60 IES come from the number of soft matrices formed by its four $V_7$-IE-loops in the reverse cascade in the opposite direction to the $V_{8A}(a_0)$-loop. From Eq. (17), a soft matrix can be obtained by adding $8\Delta_{IE}(\tau_8)$ of 8-IE energy to one $V_7$ fields. Thus, $L_{cas}\Delta_{IE}(\tau_8)/8 = H/8$ can represent the number of $V_8$-loops to perform cascading, they contained in $V_8(a_0)$-loop. Therefore,

$$H = 60\Delta_{IE}(\tau_8) - H/8 \tag{18}$$

$$H = (20/3) \cdot 8\Delta_{IE}(\tau_8) = 20/3\,\varepsilon_0(\tau_8) \tag{19}$$

And $kT_g = 20/3\,\varepsilon_0(\tau_8)$ [see Eq. (41)]; for flexible polymer chains, $kT_m = 20/3\,\varepsilon_0(\tau_8) + 4\varepsilon_0(\tau_8) = 32/3\,\varepsilon_0(\tau_8)$. Then,

$$T_g/T_m = 5/8 \quad \text{(for flexible chains)} \tag{20}$$

### 4. Free volume image, heterogeneity image and thermodynamic image

The five vacancy volumes in the center of the soft matrix in Fig. 9 correspond to the free volume theory. Since there are five local field correlations, the $b_0$, $c_0$, $d_0$ and $e_0$ vacancy volumes in the center of the soft matrix of the $A_0$ field are related to the direction and synchronization of the soft matrix in the $B_0$, $C_0$, $D_0$ and $E_0$ fields, respectively. For the average of all soft matrices, there are five vacancy volumes per 200 HSMs, resulting in a free volume fraction of 0.025. The numbers 7, 6, 5 ... 2 and 1 in Fig. 9 indicate that the IE spins of the HSM are $S_7$, $S_6$, $S_5$ ... $S_2$ and $S_1$, reflecting the heterogeneity of the system. The thermodynamic properties of the glass transition are essentially derived from the global nature of the sequential CEP excited states forming the soft matrix. The clusters can move only after waiting for the 320th IE to form the soft matrix.

### D. Modified Edwards Tube Model

#### 1. Combination of Edwards Tube Model and Reptation Model in Glass Transition

Based on the following four points, a modified Edwards's $z$-component tube model was established. (i) The three spatial component chains $N_x$, $N_y$ and $N_z$ with chain length $N$ are all the self-avoiding random walking chains in $x$, $y$ and $z$-space respectively. (ii) Only the creation - disappearance of the $\mu$-axis soft matrix can make its center HSM move along the $\mu$-axis. At this time, the central HSM is $\mu$-axis spin $S_9$, that is, the $\mu$-axis IEs surround the HSM 9 closed loops. (iii) In the glass transition, each HSM walking in the $z$-component chain is surrounded by its spin $z$-$S_9$. (iv) All possible $z$-axis interactions between polymer chains have been represented by 201 $z$-axis HSCLs in the soft matrix. The modified Edwards tube is as follows. (i) The tube fluctuations have been replaced by the generation and disappearance of $N$ $z$-axis soft matrices one after another: (ii) In the glass transition, the reptation chain is the $z$-component chain $N_x$ moving along the $+z$-direction on the center line of the tube, and the $j$-th HSM $a_j$ in the chain is the center HSM of the $j$-th soft matrix, written as $+z$-$V_8(a_j)$. (iii) The appearance of $+z$-$V_8(a_j)$ is the collective contribution of the $320N$ IEs in the fluctuating Edwards $z$-component tube. (iv) Soft matrix $z$-$V_8(a_j)$ appears at the local time $t_{8z}(a_j)$ of $a_j$ [$t_{8z}(a_j)$, $t_{8x}(a_j)$, and $t_{8y}(a_j)$ are the time when $a_j$ generates 3 $V_8$ soft matrices along the $z$, $x$, and $y$ axes respectively]. $z$-$V_8(a_j)$ disappears at time $t_{9z}(a_j)$, which will cause the vacancy volume $O(a_j)$ shared by the $N_z$ chain particles to appear in the center of the $z$-$V_8(a_j)$ at time $t_{9z}(a_j)$, and $a_j$ occupies the fraction of $O(a_j)$ (the number of DoF occupied by $a_j$) is $n_z$. Thus, at a discrete set of time points: $t_{9z}(a_0)$, $t_{9z}(a_1)$, $t_{9z}(a_2)$ ...... $t_{9z}(a_{N-1})$, the $N$ $z$-axis soft matrices complete $N$ $n_z$-step walks, the DoF number $N_z^*$ required to move the entire $N_z$ chain along the $z$-axis by $n_z$ steps: $N_z^* = N\,n_z$.

#### 2. Edwards three-component tube model in molten state

In the molten state, the improved Edward tube centerline has three self-avoiding random walking component chains. $N_z$, $N_x$, and $N_y$ [Fig. 5(b)], All the micro-cubic lattices (side length $2\Delta d$) in the tube have the same spatial orientation [Fig. 4(d)]; and each $M^+$-P in each component selects 4 faces with the same closed path among the 6 faces of the micro-cubic lattice. At time $t_{8z}(a_j)$, $t_{8x}(a_j)$, and $t_{8y}(a_j)$, $a_j$ establishes its three soft matrices on the $z$-axis, $x$-axis and $y$-axis in turn [note that the 200 HSMs in $z$-$V_8(a_j)$ are not 200 HSMs in $y$-$V_8(a_j)$ or $x$-$V_8(a_j)$]. That is, the $x$-$V_8(a_j)$ and $y$-$V_8(a_j)$ soft matrices have appeared before the time $t_{9z}(a_j)$ or before the soft matrix $z$-$V_8(a_j)$ disappears. When $z$-$V_8(a_j)$ walks $n_z$ steps, $x$-$V_8$ ($a_j$) also starts to walk along the $x$-axis. In order to make $a_j$ independently walk $n_z$ steps along the $x$-axis without being affected by the soft matrix $y$-$V_8$ ($a_j$), $a_j$ needs to consume the DoF number $N_y^*$ of the $n_y$ (=$n_z$) steps of the entire $N_y$ chain walking along the $y$-axis. Therefore, in order for $a_j$ to obtain $n_z$ DoF along the $z$-axis, $a_j$ must additionally obtain the total DoF number $N_x^* \cdot N_y^*$ of all HSMs independently walking $n_z$ steps in the $x$-$y$ space. Therefore, the number of DoF required to completely freely diffuse a reference chain $N$ in 3D space is $N^* = N_x^* N_y^* N_z^*$.

### E. Anomalous Relaxation Law of Glass State

The abnormal relaxation of the glass state has been a research topic of concern in the theoretical world. Generally, the physical quantity $\xi(t)$ in a system will return according to the physical law of Eq. (21) when the system deviates from its equilibrium state.

$$\xi(t) \sim \exp[-(t/\tau)] \qquad (21)$$

However, if the glassy system is driven out of equilibrium, it returns according to the formula (22)

$$\xi(t) \sim \exp[-(t/\tau)^\beta] \qquad (22)$$

Physicists have been troubled by this unusual relaxation, and mathematical models of glass states proposed by many scholars are also based on this experimental law. Now we try a new perspective of soft matrix with CEP excited states to directly prove Eq. (22).

Since inverse cascade–cascade motions only occur in some discrete "lakes" in glass state when $T < T_g$, Eq. (21) is still holds true in the "lake" regions as long as the $t$ in Eq. (21) is the local domain time. One of the key concepts is that the equilibrium state of glass state is the equilibrium state between the random thermo-vibration energy $kT$ and the energy of all $V_8$-IE closed-loops. The number of $V_8$-IE closed-loops always dominates the number of the excited domains (lakes) at temperature $T$. That is, when $T < T_g$, the soft matrix composed of $V_8$-IE loops in the system is not connected into one piece. The temperature increase only increases the number density of the $V_8$-IE loops in the system (only at $T_g$ temperature, the dynamic $V_8$ cluster soft matrices are connected to infinity).

Suppose that the glass state we observe is an unbalanced state, which comes from an equilibrium state at temperature $T_1$, and suddenly drops to temperature $T_2$ at times $t = 0$ and $T_2 < T_1 < T_g$. During the relaxation time of $t$, the entire relaxation energy of these $V_8$-IE closed-loops is $\varepsilon = k(T_1 - T_2) = k\Delta T$. From the famous Kolmogorov law in cascade [63,64]:

$$l_i^2 / t_i^3 = k\Delta T/t = \text{constant} \qquad (23)$$

Where $l_i$ is the diameter length of 2D $V_i$-cluster and $k\Delta T/t$ is the cascade energy mobility; $t_i$ is the local domain time and $t$ is the relaxation time in system. The cascade of $V_8$-IE loop in each "lake" satisfies Eq. (23). Therefore, on average, the relaxation cascades of all $V_8$-IE-loops in different directions in the system are expressed by Eq. (23). From Eq. (23)

$$t_i = \left(\frac{l_i^2 t}{k\Delta T}\right)^{\frac{1}{3}} \qquad (24)$$

Substituting the local domain time $t_i$ in Eq. (24) into Eq. (21)

$$\xi(t) = \exp(-\frac{l_i^2 t}{k\Delta T \tau_{loc}^3})^{\frac{1}{3}} = \exp[-(\frac{t}{\tau_{sys}})^\beta] \qquad (25)$$

It can be seen that the general physical relaxation theorem, Eq. (22), is still valid on the domain scale in the glass state. The abnormal mathematical expression Eq. (25) in the glass state is only the *emergent property* of domains in system.

### F. Theoretical Proof of the Empirical WLF Equation in Glass Transition

The discovery of the standard Willams-Landel-Ferry (WLF) empirical equation (26) has been nearly 70 years old, but so far, there is no theory that can strictly prove the WLF equation. In order to verify the IE soft matrix theory, one can try to use it to directly prove the WLF equation. The theoretical proof shows that the constant $c_1$ in the equation, taking logarithm, is the (dimensionless) cooperative orientation activation energy ($\Delta E_{co}$), and $c_2 k_B$ the (dimensionless) potential well energy $\varepsilon_0$ ($\tau_8$). The well-known Clapeyron equation for controlling first-order phase transitions in thermodynamics is only applicable to each newly generated "transient phase change" system that occurs when $\Delta T$ temperature is continuously added to the system. In each of a number of newly occurring transient systems in which the temperature $\Delta T$ is continuously increased, the WLF equation is derived by repeatedly applying a first-order phase transformation law. The well-known semi-empirical WLF equation

$$\log \frac{\eta(T)}{\eta_g} = -\frac{C_1(T-T_g)}{T-T_g+C_2} = -\frac{17.44(T-T_g)}{T-T_g+51.6} \qquad (26)$$

In which $c_1$ and $c_2$ are two experimental constants for most flexible polymer. When $\Delta T$ temperature is continuously added to the temperature $T$ system, applying the Clapeyron Equation:

$$\Delta\sigma(T) \cdot \Delta V(T) = -\Delta T \cdot \Delta S(T) \qquad (27)$$

Assign a negative sign to the right side of Eq. (27) because here $\Delta\sigma$ is an increment of tensile stress of $\sigma$, not the increase in compressive stress in liquid-gas phase transition. $\Delta S(T)$ is the entropy change caused by the increase in the number of soft matrices in the system when the temperature $T$ rises to the temperature $T + \Delta T$. Although both the $T$ system and the $(T + \Delta T)$ system are ideal randomly distributed systems, the latter will have more ordered soft matrices embedded than the former, and the sum of the energies of these soft matrices is $k(\Delta T)$.

When the temperature is $T_k$, the system will undergo the percolation transition of $V_7$-$V_6$ clusters (Fig. 8), and the orderly IE energy of all $V_7$-$V_6$ clusters is in equilibrium with the disordered energy of the zero entropy temperature of $kT_K$. When all $V_8$–$V_7$ are dynamically connected, the $V_8$ soft matrix system corresponds to the glass transition at temperature $T_g$. As the temperature $T$ continues to increase, the number of $V_8$ soft matrices continues to increase in the $V_7$-$V_6$ percolation field with $kT_K$ (=$kT_2$) disorder energy.

Thus, the ordered energy of all the soft matrices in the $T$-temperature system is balanced with the disordered energy $k(T-T_2)$. The $z$-axis external stress $\sigma(T)$ should be balanced with the "conformational entropy stress of the Edwards $z$-component tube" excited in the system (the conformational entropy stress formed along the $z$-axis of all randomly oriented $V_8$ soft matrices in the system), and the volume change $\Delta V(T)$ in the system at temperature $T$ corresponds to the total volume of the free volumes in the tensile sample, that is

$$\sigma(T) \cdot \Delta V(T) = k(T - T_2) \quad (28)$$

The energy of the 320 IEs surrounded by 200 HSMs is the cooperative orientation activation energy of soft matrix, from Eq. (16)

$$\Delta E_{co} = 320 \Delta_{IE} = 40\varepsilon_0 \quad (29)$$

$T_2$ is the temperature of the clusters $V_7$-$V_6$ percolation transition in the system, and is also the temperature at which the entropy change associated with the soft matrix jumping (not the molecular jumping) becomes zero, that is

$$\Delta S(T) = \Delta E_{co} / (T - T_2) \quad (30)$$

From Eqs (27-30)

$$\Delta \sigma(T) / \sigma(T) = -\Delta E_{co} \cdot \Delta T / k (T - T_2)^2 \quad (31)$$

From $k_B T_g = k_B T_2 + \varepsilon_0$, Eq. (29) is rewritten as

$$\Delta\sigma(T)/\sigma(T) = -\Delta E_{co} \cdot \Delta T / \{k_B [T - \frac{(k_B T_g - \varepsilon_0)}{k_B}]^2\} \quad (32)$$

In experiments at constant rate extensional viscosity, $\Delta \sigma(T)/\sigma(T) = \Delta \eta(T)/\eta(T)$, we have

$$\int \frac{d\eta(T)}{\eta(T)} = -\frac{\Delta E_{co}}{k} \int_{T_g}^{T} \frac{dT}{\left(T - \frac{kT_g - \varepsilon_0}{k}\right)^2}$$

$$\ln \frac{\eta(T)}{\eta_g} = -\frac{\Delta E_{co}}{\varepsilon_0} \cdot \frac{T - T_g}{T - T_g + \frac{\varepsilon_0}{k}} = -\frac{40(T - T_g)}{T - T_g + \frac{\varepsilon_0}{k}}$$

$$\log \frac{\eta(T)}{\eta_g} = -\frac{17.37(T - T_g)}{T - T_g + \frac{\varepsilon_0}{k}} \quad (33)$$

Eq. (33) is in the form of the standard WLF equation. $k C_2$ is the potential well energy $\varepsilon_0$, $\varepsilon_0 = 51.6k$. And Eq. (33) derives the famous relationship: $\eta \sim \exp[A/(T - T^\star)]$, where $T^\star$ is called a "ghost" transition by de Gennes [6]. From Eqs (16), (17), the average CEP interface excitation energy

$$\Delta_{IE}(\tau_8) = 1/8 \times 51.6\ k = 6.45k \approx 0.56 \text{ meV} \quad (34)$$

### G. Neighborhood Effect and Thermodynamic Second Virial Coefficient

The neighborhood effect only considers the interaction between two adjacent HSMs, so as to obtain the largest 2D ordered soft matrix energy $kT_g°$ (= $kT_g$) against thermal random vibration in the entire solid-liquid transition temperature range. The two-body interaction in the neighborhood effect should have the same energy $kT_g°$ as the two-body interaction represented by the thermodynamic second virial coefficient. Therefore, it is meaningful to discuss the relationship between these two two-body interactions, at least to make people clarify the role of the "free volume concept" in the two types of statistics. During the solid-liquid transition, the rare soft matrix central vacancy cavity $O(a_j)$ in the system can be regarded as an "ideal gas molecule", The expansion of the virial coefficient in thermodynamics is

$$\frac{PV}{kT} = 1 + B_2 + B_3 + ... \quad (35)$$

Where $B_2$ and $B_3$ are the reduced second- and third-virial coefficients, respectively. It can be strictly proved that the reduced third virial coefficient for hard-sphere system is constant, $B_3 \equiv 5/8$, [65, 66] independent of temperature and cluster volume. During the glass transition, the volume changes abnormally, the partial derivative of $B_2$ to the volume variable $V$ is obtained from Eq. (35).

$$\frac{\partial B_2}{\partial V} = \frac{P}{kT} = \frac{B_2}{V} \quad (36)$$

In statistical mechanics, the abnormal thermal capacity $C_p$ occurs in self-similar system. From *enthalpy* $H = E + PV$, definition of Joule-Thomson coefficient [66] $\mu_j$ is

$$\mu_J = (\partial T/\partial P)_{H,N} = C_P^{-1}\left[T(\partial V/\partial T)_{P,N} - V\right] \quad (37)$$

In the glass transition, $C_p$ is abnormal, so, $\mu_j \equiv 0$ can correspond to the glass transition. When $\mu_J \equiv 0$

$$(\partial V/\partial T)_{P,N} \equiv V/T \quad (38)$$

Substituting Eq. (38) into Eq. (36) to obtain the relationship between the second virial coefficient and temperature under abnormal conditions of glass transition heat capacity.

$$\frac{\partial B_2}{k\partial T} = \frac{B_2}{kT} \quad (39)$$

For L−J potential hard-sphere model, the second virial coefficient may have a parsed expression [67]:

$$B_2(T^*)_{L-J} \equiv B_2(T^*)_{L-J} / b_0$$

$$= \frac{4}{T^*} \int_0^\infty dx \cdot x^2 \left[\frac{12}{x^{12}} - \frac{6}{x^6}\right] \exp\left\{-\frac{4}{T^*}\left[\left(\frac{1}{x}\right)^{12} - \left(\frac{1}{x}\right)^6\right]\right\}$$

$$= \sum_{n=0}^{\infty} \alpha_n \left(\frac{1}{T^*}\right)^{\frac{2n+1}{4}} \quad (40)$$

Where $\alpha_n$ can be represented by the $\Gamma$ function:

$$\alpha_n = -\frac{\sqrt{2}\,\Gamma\left(\frac{2n-1}{4}\right)}{2^{(2-n)} n!}$$

In Eq. (40), $x = q/\sigma$, $q$ and $\sigma$ are the $q$-axis distance and the hard-sphere diameter in the L–J potential, $kT/\varepsilon_0$ is the reduced temperature $T^*$, $\varepsilon_0$ is the potential well energy of the L–J potential, and $k$ is the Boltzmann constant. Note that $x$ in Eq. (40) uses a dimensionless ($q/\sigma$) ratio, similar to Eq. (2), and the temperature $T$ is measured in $\varepsilon_0$ as a unit 1.

Mathematical Eq. (40) shows the structural complexity of amorphous materials. Substituting Eq. (40) into Eq. (39), the resulting equation should contain a lot of structural information of the glass state and glass transition, but unfortunately, no analytical solution can be obtained. It can be used as a graph method to find a unique set of approximate solutions, see Fig. 10. Graphical values are referenced from the literature [67].

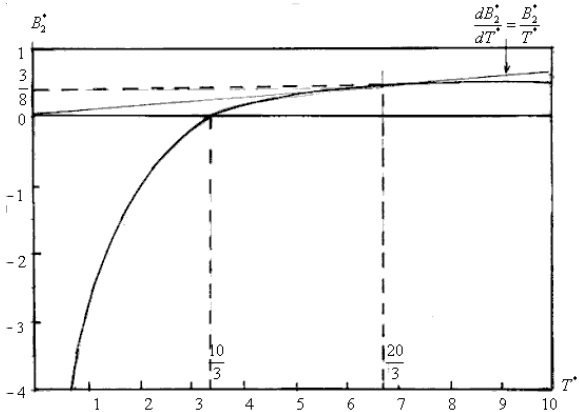

FIG. 10. Graphic solution results of Eqs (39)-(40)

The only set of approximate solutions in Fig10 are.

$B_2^* \approx 3/8$, $T^* \approx 20/3$, or, $kT(B_2^*) \approx 20/3\ \varepsilon_0$
(41)

Eqs (36–41) give the following physical image. At the glass transition temperature $T_g$, for an "ideal gas" system with a constant number of vacancies in each local volume of the system (free volume fraction 0.025), when the external pressure is constant, Eq. (38) holds, and the physical quantities on each small area of the system $B_2/\partial T$ and $\partial V/\partial T$ are the same as the physical quantities $B_2/T$ and $V/T$ of the system. The IE energy required to meet the conditions of "ideal gas molecule" can be provided by $C_p$ in Eq. (37). At this time, the reduced temperature of the system is $T^* \approx 20/3$, which is the same as the reduced energy used to generate the soft matrix in Eq. (19) (also represented by the reduced temperature). Comparing Eq. (41) and Eq. (19), we can see that the Hamiltonian describing the excited states of CEPs is the energy of the glass transition temperature of the system. After $T_g$, as the temperature continues to rise, more and more nanoscale small areas will continue to undergo an orderly 2D soft matrix transformation, thereby keeping the thermal expansion rate [Eq. (38)] of the system unchanged. This also explains why the energy $kT_g°$ appears in systems where the temperature is higher than $T_g$ or even higher than $T_m$.

The theory of two-body interaction in thermodynamics provides a way for the emergence of the CEP excited states in the neighborhood effect. That is, without violating the theory of thermodynamics, the coupled electron pair in the neighborhood effect escapes from the two C–H bond resonance states to the overlapping interface of the two HSMs, thereby forming the CEP excited state interface, HSM IE spin, and exclude volume interface, and the required additional IE energy appears in the form of abnormal heat capacity $C_p$.